\documentclass[appendixfloats, numberedappendix]{emulateapj}

\usepackage[utf8]{inputenc}
\usepackage[usenames,dvipsnames]{color}
\usepackage{natbib}
\usepackage{graphicx}
\usepackage{subfigure}
\usepackage{comment}
\usepackage{amsbsy}
\usepackage{amsmath}
\usepackage{float}
\usepackage{amsmath}
\usepackage{fourier}
\usepackage{xfrac}
\usepackage{mathtools}
\usepackage{booktabs}
\usepackage{array}
\usepackage{enumitem}

\begin{document}
\title{Broadband Intensity Tomography:\\Spectral Tagging of the Cosmic UV Background}
\shorttitle{UV Background}
\shortauthors{Chiang, M{\'e}nard, \& Schiminovich}

\author{Yi-Kuan Chiang\altaffilmark{1}, Brice M{\'e}nard\altaffilmark{1,2}, David Schiminovich\altaffilmark{3}}
\altaffiltext{1}{Department of Physics \& Astronomy, Johns Hopkins University, 3400 N. Charles Street, Baltimore, MD 21218, USA}
\altaffiltext{2}{Institute for the Physics and Mathematics of the Universe, Tokyo University, Kashiwa 277-8583, Japan}
\altaffiltext{3}{Department of Astronomy, Columbia University, New York, NY 10027, USA}

\submitted{Accepted for publication in the Astrophysical Journal (ApJ)}

\begin{abstract}
Cosmic photons can be efficiently collected by broadband intensity mapping but information on their emission redshift and frequency is largely lost. We introduce a technique to statistically recover these otherwise collapsed dimensions by exploiting information in spatial fluctuations and apply it to the Galaxy Evolution Explorer (GALEX) All Sky and Medium Imaging Surveys. By spatially cross-correlating photons in the GALEX far-UV $(1500\,\rm \AA)$ and near-UV $(2300\,\rm \AA)$ bands with a million spectroscopic objects in the Sloan Digital Sky Survey as a function of redshift, we robustly detect the redshift-dependent intensity of the UV background (UVB) modulated by its clustering bias up to $z\sim2$. These measurements clearly reveal the imprints of UVB spectral features redshifting through the filters. Using a simple parameterization, we simultaneously fit a UVB emissivity and clustering bias factor to these observations and constrain the main spectral features of the UVB spectrum: (i) the Lyman break, (ii) the non-ionizing UV continuum, which agrees with the Haardt \& Madau model but does not rely on any assumption regarding the nature of the sources, and (iii) the Ly$\alpha$ emission, whose luminosity density is consistent with estimates of the combined galaxy and AGN contributions at $z\sim1$. Because the technique probes the total background including low surface brightness emission, we place constraints on the amount of UV light originating from the diffuse intergalactic medium (IGM). Finally, the clustering bias of UV photons is found to be chromatic and evolving. Our frequency- and redshift-dependent UVB measurement delivers a summary statistic of the universe's net radiation output from stars, black holes, and the IGM combined.
\end{abstract}

\section{Introduction}\label{sec:intro}
Photons in the extragalactic background light (EBL) can be more efficiently collected in broadband observations than in spectroscopic observations, but information on their emission redshift and frequency gets diluted or lost. Depending on the instrument and survey depth, a significant fraction of the collected photons may belong to a diffuse component outside detected sources. Such photons are often discarded, together with potentially valuable astronomical information. For this reason, intensity mapping is being developed as a technique to measure and analyze the total radiation as a continuous field as opposed to the study of discrete objects. It provides us with a powerful probe of the universe that does not rely on the use of a surface brightness thresholding required for source detection. Using this approach, the study of the three-dimensional universe can be enabled by targeting specific emission lines and selecting redshifts by tuning the frequency of the observations. This is referred to as line intensity mapping, and a recent review on the subject is given by \cite{2017arXiv170909066K}.

Being able to use the concept of intensity mapping with broadband data in a redshift-dependent manner would open up a number of new scientific explorations. In this work, we develop a new method to statistically tag the rest-frame frequencies of EBL photons (in a diffuse field and/or detected sources) in broadband observations with a spectral resolution finer than that of the bandwidth. This is achieved by combining the technique of clustering-based redshift inference \citep{2008ApJ...684...88N,2013arXiv1303.4722M} and a data-driven estimation of the long-established concept of the $K$-correction \citep{1956AJ.....61...97H,2002astro.ph.10394H}. We can measure the cosmic $K$-correction, i.e., the differential EBL intensity as a function of redshift using the clustering technique. Because this $K$-correction depends on the spectral energy distribution of the EBL, one can constrain the main spectral features in the EBL.

We apply this technique to study the cosmic ultraviolet background (UVB) in the All Sky and Medium Imaging Surveys of the Galaxy Evolution Explorer (GALEX) satellite. Astrophysically, the UVB is of critical importance as the photoionization and excitation of most of the atomic elements are tied to this radiation field. The overall amplitude and redshift evolution of the UVB traces galaxy formation and the cosmic star-formation history \citep{2014ARA&A..52..415M}. The Lyman--Werner background in the UV photodissociates molecular hydrogen and regulates galaxies' star-formation efficiency especially in the early universe \citep{1997ApJ...476..458H}. The metagalactic UVB also provides a starting point in modeling the circumgalactic medium (CGM) in both absorption \citep{2014ApJ...792....8W} and emission \citep{2016ApJ...827..148C}. Finally, the diffuse intergalactic medium (IGM) is expected to radiate in the UV in both the continuum and Ly$\alpha$ \citep{1974Natur.247..513D,1980Natur.288..119P,2012ApJ...746..125H}; only very recently have observational studies started to deliver the first detections, but they are still limited in fluorescent radiation near bright quasars \citep{2014Natur.506...63C,2014ApJ...786..107M}.

In this paper, we present new constraints on the spectrum of the UVB volume emissivity at $z<2$ based on GALEX imaging data. This extends the explorations of intensity mapping from radio \citep{2010Natur.466..463C}, infrared \citep{2018MNRAS.478.1911P}, optical \citep{2011MNRAS.417..801M,2016MNRAS.457.3541C,2018MNRAS.481.1320C}, to the ultraviolet. Throughout this paper, we assume the \cite{2014A&A...571A..16P} cosmology. All of the cosmic volumetric quantities are expressed in comoving units. Magnitudes are given in the AB system.

\section{The Framework}\label{sec:method}

We introduce a broadband intensity tomography technique to recover the spectrum of the universe and its evolution over cosmic time. The target quantity is thus a frequency- and time-dependent EBL volume emissivity. The general approach is to \textbf{(i)} deproject the observed broadband EBL intensity into differential contributions as a function of redshift using the technique of clustering-based redshift estimation and \textbf{(ii)} build a generative forward model describing the observed redshift deprojected intensity given any input EBL emissivity and fit to the data under the Bayesian framework to determine the emissivity posterior distribution. This framework allows us to propagate the information content from spatial fluctuations in the broadband intensity maps to the redshift distribution of EBL photons, and finally to its spectral features and cosmic time evolution. It allows us to address some science questions that are typically thought to be accessible only with spectroscopic intensity mapping data. It is applicable to all wavebands across the electromagnetic spectrum. It can be viewed as a generalization of line intensity mapping to arbitrary spectral features in data with arbitrary bandwidth, with the line-of-sight projection or line confusion \citep[e.g., ][]{2016ApJ...832..165C} solved by clustering redshift tomography. Below we describe our technique in detail.

\subsection{Information content in redshift}\label{subsec:formalism}

The redshift $z$ of a photon, by definition, carries spectral information that quantifies the fractional change in its frequency or wavelength between the emitted and observed frames. On cosmological scales, the expansion of the universe relates redshift to the distance or cosmic time at which the photon was emitted. Ignoring the effects of peculiar velocities, the information content carried by any redshift-dependent quantity $X=X(z)$ is
\begin{eqnarray}
\Bigg(\frac{\textrm{d}}{\textrm{d} z}\Bigg)\, X &=& 
\Bigg(\frac{\textrm{d} \nu}{\textrm{d}z}\frac{\partial}{\partial \nu} + \frac{\textrm{d}t}{\textrm{d}z}\frac{\partial}{\partial t}\Bigg)\, X \;,
    \label{eq:ddz}
\end{eqnarray}
where $\nu$ is the frequency of the photon and $t$ is the cosmic time. Equation~\ref{eq:ddz} implies that by constraining the redshift dependence of the quantity of interest, we can also probe its frequency and time dependence. Conversely, this also presents the challenge of breaking the potential degeneracies between these two quantities. Most of the time, Equation~\ref{eq:ddz} is used to infer the $\partial/\partial t$ term, with the $\partial/\partial \nu$ term modeled out via the so-called ``$K$-correction'' \citep{1956AJ.....61...97H,2002astro.ph.10394H}. One of our major goals here is to explore the full power of Equation~\ref{eq:ddz} without making strong assumptions on the $\partial/\partial \nu$ term: by making direct measurements of the redshift dependence of the quantity $X$, can we obtain meaningful constraints on the ``spectrum'' of $X$?

\subsection{EBL integral constraint}\label{subsec:cosmoRT}

Our physical quantity of interest is the spatially averaged, metagalactic comoving emissivity $\epsilon_{\nu} = \epsilon_{\nu}(\nu, z)$\footnote{We follow the convention that, depending on the context, the $z$ label sometimes refers to solely a $t$ label as on the right-hand side in this expression.} of the EBL (in units of $\rm erg\; s^{-1}\; Hz^{-1}\; Mpc^{-3}$), as it provides a summary statistic of the total radiation output in the universe as a function of frequency and cosmic time. Observationally, one measures the specific intensity $j_{\nu}$ ($\rm erg\; s^{-1}\; cm^{-2}\; Hz^{-1}\; sr^{-1}$). In the expanding universe, these two can be related by the cosmological radiative transfer equation \citep[e.g.,][]{1997ApJ...486..581G}:
\begin{eqnarray}
\Bigg(\frac{\partial}{\partial t} - \nu\, H\,\frac{\partial}{\partial \nu}\Bigg)\, j_{\nu} + 3H\, j_{\nu} = -\textrm{c}\,\kappa\,j_{\nu}  
+ \frac{\textrm{c}}{4 \pi}\, \epsilon_{\nu}\,(1+z)^3\;,
    \label{eq:radiative_transfer}
\end{eqnarray}
where $\textrm{c}$ is the speed of light, $H$ is the Hubble parameter, $\epsilon_{\nu}$ serves as the source term, and the opacity $\kappa$ characterizes the sink term due to IGM absorption. Integrating over the entire line-of-sight path length, or equivalently, over redshift, we get
\begin{eqnarray}
j_{\nu_{\rm obs}}(\nu_{\rm obs}) = \frac{\textrm{c}}{4\pi}\, \int^{\infty}_0\, \textrm{d}z\, \left|\frac{\textrm{d}t}{\textrm{d}z}\right|\, \epsilon_{\nu}(\nu, z)\, e^{-\tau}
\label{eq:J_nu}
\end{eqnarray}
for an observer at $z=0$, where $\nu_{\rm obs}$ is the observed-frame frequency, $\nu = \nu_{\rm obs}\; (1+z)$,  $\left|\textrm{d}t/\textrm{d}z\right| = H(z)^{-1}(1+z)^{-1}$ and $\tau = \int \kappa\,\textrm{d}s$ is the optical depth describing IGM absorption along the line of sight, with $\textrm{d}s$ being the path-length element. Photometric observations using a filter $i$ for which the response is denoted by $R^i$ lead to a band-averaged specific intensity,
\begin{eqnarray}
J_{\nu_{\rm obs}}^{i} = \int \frac{\textrm{d} \nu_{\rm obs}}{\nu_{\rm obs}}\, j_{\nu_{\rm obs}}(\nu_{\rm obs})\, R^i(\nu_{\rm obs})\;,
\label{eq:J_nu_Broadband}
\end{eqnarray}
where $R^i(\nu)$ is proportional to the number of electrons yielded per incident photon (i.e., QE, appropriate for photon-counting detectors) and is normalized such that $J_{\nu} = j_{\nu}$ for a flat $j_{\nu}$. The quantity $J_{\nu}^i$ is an integral constraint of $\epsilon_{\nu}$ collapsed over cosmic time and a range of frequency.

\subsection{Redshift tomography using clustering}\label{subsec:redshift_tomo}
Both the EBL emissivity and its integral, the intensity, are generally spatially varying; they trace the underlying matter density field in 3D and 2D, respectively. In a broadband intensity map of a certain angular resolution, $J_{\nu} =  J_{\nu}(\phi)$, the phase information of the corresponding angular fluctuations actually carries redshift information if we know the mapping between the two. Such a mapping can be obtained using another set of matter tracers, usually bright objects in spectroscopic redshift surveys over the same patch of the sky and propagated into the intensity field via cross-correlations. This provides a way to recover the redshift distribution of photons in the map. This so-called clustering redshift technique is laid out in \cite{2008ApJ...684...88N}, \cite{2013MNRAS.433.2857M}, and \cite{2013arXiv1303.4722M}, and tested against simulations in \cite{2010ApJ...721..456M} and \cite{2013MNRAS.431.3307S}. It has been applied to a wide range of survey datasets to estimate the redshift probability distributions of discrete objects \citep{2013arXiv1303.4722M, 2015MNRAS.447.3500R, 2016MNRAS.457.3912R, 2016MNRAS.460..163R, 2016MNRAS.462.1683S, 2017MNRAS.467.3576M, 2018MNRAS.477.2196D} as well as that of the diffuse radiation field \citep{2015MNRAS.446.2696S,2019ApJ...870..120C}. We refer the readers to these papers for details. Here we briefly describe the formalism tailored for broadband tomography.

The goal here is to measure the global emission redshift distribution for the EBL intensity as seen for a redshift $0$ observer. This can be expressed as $\textrm{d} J_{\nu}/\textrm{d}z(z)$ (same units as $J_{\nu}$), whose integral makes up the total observed projected intensity,

\begin{eqnarray}
J_{\nu} = \int_0^{\infty} \frac{\textrm{d}J_{\nu}}{\textrm{d}z}(z)\,\textrm{d}z\;.
\label{eq:dJdz_normalization}
\end{eqnarray}
On the map level, we make distinctions between the local and spatially averaged intensity and define an overdensity field 
\begin{eqnarray}
    \Delta J_{\nu}(\phi) &=&  J_{\nu}(\phi) - \langle J_{\nu} \rangle\,,
    \label{eq:delta_J}
\end{eqnarray}
where $\langle\, \rangle$ denotes the ensemble or large-scale average. Here we choose to use the absolute instead of a fractional overdensity field to reduce the impact of Galactic or sky foregrounds in the analysis, which will become clear later. To measure $\textrm{d} J_{\nu}/\textrm{d}z(z)$ we can make use of an external set of reference galaxies or quasars, whose redshifts are already known spectroscopically. The corresponding overdensity field can be written in angular plus redshift ($2+1$D) space,
\begin{eqnarray}
    \delta_r(\phi, z) &=& \frac{{\rm n}(\phi,z)-\langle {\rm n} (z)\rangle}{\langle {\rm n} (z)\rangle}\,,
    \label{eq:delta_r}
\end{eqnarray}
where $n$ is the number density of the reference objects. Under a linear assumption, the reference traces the underlying matter density $\delta_m$ (defined analogously to $\delta_r$ in Equation~\ref{eq:delta_r}) with a linear bias factor $b_r$ such that $\delta_r = b_r \delta_m$.

With its 3D information, the reference sample provides a mapping between the phases of the projected spatial fluctuations and redshift. To extract this mapping, we can measure the angular cross-correlation between the intensity field and the reference as a function of the redshift of the latter,
\begin{eqnarray}
w_{Jr}(\theta, z) &=&  \langle \Delta J_{\nu}(\phi)\, \cdot \,\delta_r(\phi+\theta, z) \rangle \nonumber \\
&=& \langle J_{\nu}(\theta, z)\rangle_r -  \langle J_{\nu} \rangle\;,
\label{eq:w_theta}
\end{eqnarray} 
where $\langle J_{\nu}(\theta, z)\rangle_r$ denotes the mean intensity at angular separation $\theta$ around reference objects at a given redshift. This estimator using the absolute intensity overdensity is ideal for extragalactic intensity data, whose normalization is hard to estimate due to the presence of a considerable foreground. As long as the foreground does not correlate with extragalactic large-scale structures, its contribution is canceled out in the two terms on the right-hand side, thus the $w_{Jr}$ estimator is unbiased. The goal here is to relate $w_{Jr}(\theta, z)$ to the desired quantity $\textrm{d} J/\textrm{d}z(z)$. Because the latter does not depend on $\theta$, we can integrate $w_{Jr}$ over $\theta$ to get a one-bin measurement to enhance the signal-to-noise ratio,
\begin{eqnarray}
\bar{w}_{Jr}(z) = \int_{\theta_{\rm min}}^{\theta_{\rm max}} W(\theta)\,w_{Jr}(\theta, z)\, \textrm{d}\theta\;,
\label{eq:wbar}
\end{eqnarray}
where $W(\theta)$ is an arbitrary weight function carrying the units of $\rm \theta^{-1}$. Following \cite{2013arXiv1303.4722M}, we set a normalized $W(\theta) \propto \theta^{-0.8}$, the same angular scaling as that of the typical galaxy angular correlation functions, which optimizes the signal-to-noise ratio. We set our integration boundaries $\theta_{\rm min}$--$\theta_{\rm max}$ to those corresponding to $0.5$--$5$ physical Mpc at each redshift bin; this is chosen to avoid strongly nonlinear clustering at small scales and ab uncontrolled zero point in typical photometry datasets at large scales.

We are now ready to relate the observable $\bar{w}_{Jr}(z)$ in Equation~\ref{eq:wbar}, usually measured in bins of redshift, to the redshift decomposition $\textrm{d} J/\textrm{d}z(z)$. The $\bar{w}_{Jr}(z)$ is a light-weighted, i.e., $\textrm{d} J/\textrm{d}z$-weighted, estimator of the underlying matter autocorrelation modulated by the clustering bias of both tracers. Given a matter power spectrum $P(k, z)$, we can obtain a 2D--3D matter correlation between an infinitely thin slice of matter at redshift $z$ and the projected matter density field,
\begin{eqnarray}
w_{m}(\theta, z) &=& \frac{1}{2\pi}\,\int_0^{\infty}\,\textrm{d}k\, k\, P(k, z)\,J_0(k\,\theta\,X(z))\,\frac{\textrm{d}z}{\textrm{d}X}\, ,
\label{eq:w_m}
\end{eqnarray}
\citep{1953ApJ...117..134L,2005ApJ...619..147M}, where $J_0$ is the Bessel function of the first kind and $X(z)$ is the comoving radial distance. For $P(k,z)$, we use the nonlinear matter power spectrum calculated using the CLASS code \citep{2011arXiv1104.2932L}. The tracer cross-correlation (Equation~\ref{eq:w_theta} and \ref{eq:wbar}) then can be linked to the matter correlation as follows: 
\begin{eqnarray}
\bar{w}_{Jr}(z) = 
\Bigg( \frac{\textrm{d} J_{\nu}}{\textrm{d}z}(z)\, b_J(z) \Bigg)\,\Bigg( b_r(z)\,\bar{w}_{m}(z) \Bigg)\;,
\label{eq:wbar_to_dJdz}
\end{eqnarray}
where $\textrm{d} J/\textrm{d}z$ gives the redshift-dependent light weighting to the correlation amplitude, $b_J$ is the effective clustering bias factor of the intensity, and $\bar{w}_{m}(z)$ is $w_{m}(\theta, z)$ integrated over $\theta$ the same way as $w_{Jr}(\theta, z)$ in Equation~\ref{eq:wbar}. Equation~\ref{eq:wbar_to_dJdz} here effectively defines $b_J$, while alternatively, in the next section, we will write $b_J$ as a band-averaged version of the bias factor for the underlying EBL photon field, which could be both redshift and wavelength dependent. In Equation~\ref{eq:wbar_to_dJdz}, $\bar{w}_{m}$ is determined by the cosmology, $b_r$ can be measured using the autocorrelations of the reference objects, and our $\bar{w}_{Jr}$ estimator measured in bins of redshift thus constrains the product of $\textrm{d}J_{\nu}/\textrm{d}z(z)$ and $b_J(z)$.

\subsection{Spectral tagging}\label{subsec:spectral_tagging}

\begin{figure}[t!]
    \begin{center}
         \includegraphics[width=0.475\textwidth]{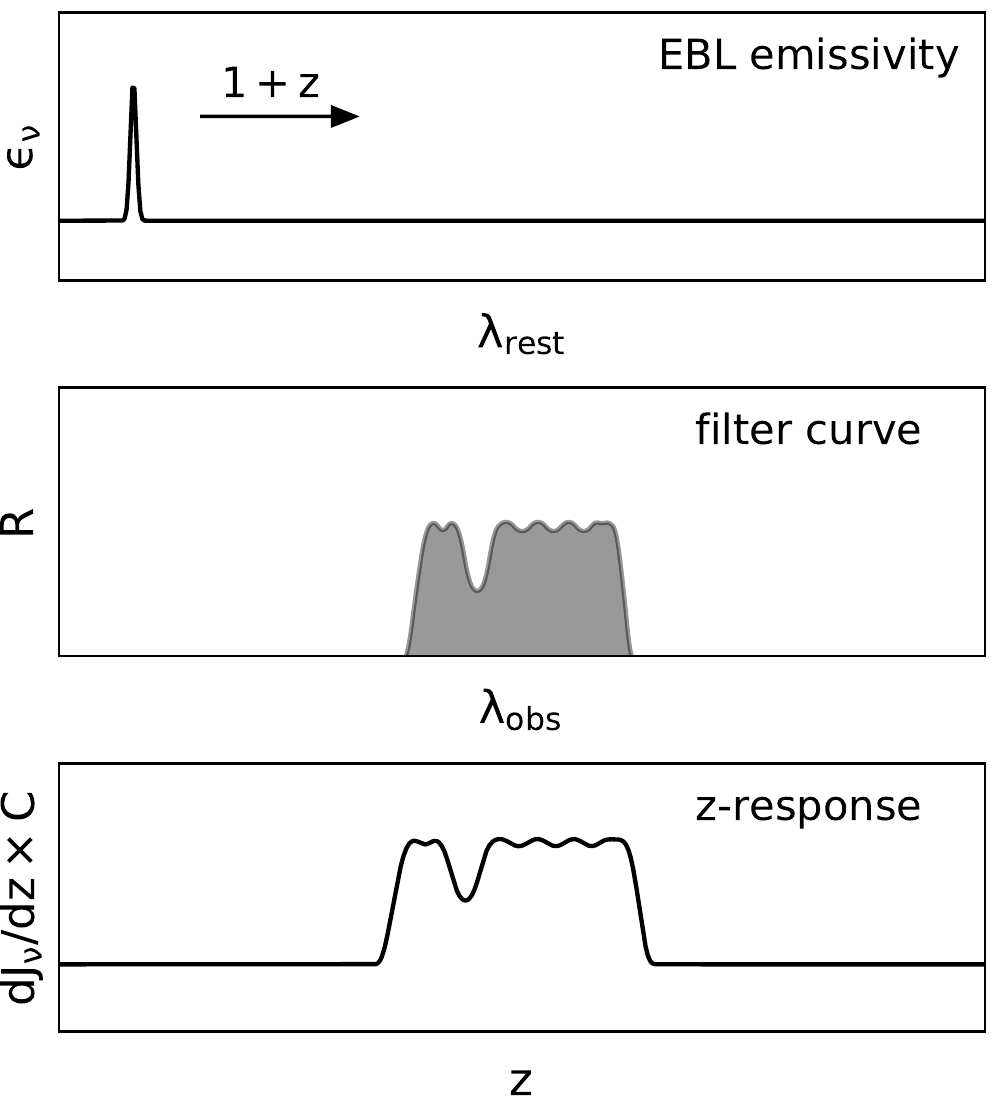}
    \end{center}
    \vspace{0.75mm}
    \caption{Spectral tagging in the simplest case. The EBL emissivity (top panel) is assumed to be a flat continuum with a line and does not evolve over cosmic time. After being observed with a broadband (middle panel) and deprojected using the clustering redshift technique, we get a redshift response (bottom panel) that simply reflects the shape of the filter curve. The factor $C \propto H(z)\, (1+z)$ in the y-axis of the bottom panel is to cancel out trivial redshift factors carrying no information under a fixed cosmology.}
    \label{fig:ideal_case}
\end{figure}

The observable $\bar{w}_{Jr}(z)$ carries information on the combination ${\textrm{d} J_{\nu}}/{\textrm{d}z}(z)\, b_J(z)$. Following Equations~\ref{eq:J_nu}--\ref{eq:dJdz_normalization}, the first term of this product is given by
\begin{eqnarray}
\frac{\textrm{d} J_{\nu_{\rm obs}}}{\textrm{d}z}(z) &=& \frac{\textrm{c}}{4\pi\,H(z)\, (1+z)}\, \int\, \frac{\textrm{d}\nu_{\rm{obs}}}{\nu_{\rm{obs}}} \, R(\nu_{\rm{obs}})\, \epsilon_{\nu}(\nu, z)\, e^{-\tau}\;,\nonumber\\
&&
\label{eq:dJdz_to_epsilon}
\end{eqnarray}
where $\nu = \nu_{\rm{obs}}\; (1+z)$. We now illustrate how we can use the observable $\bar{w}_{Jr}(z)$ to constrain the EBL emissivity $\epsilon_{\nu}(\nu, z)$.
To first gain intuition on the extraction of spectral information based on tomographic redshift measurements, let us first consider a special case of a nonevolving EBL emissivity with a single emission line, whose bias $b_J$ is also redshift independent and the line-of-sight absorption can be ignored. In this simplistic scenario, the redshift trend $\textrm{d}J_{\nu}/\textrm{d}z$ is given by a sliding integral of the spectral feature $\epsilon_{\nu}$ with the filter curve. This can be visualized in Figure~\ref{fig:ideal_case}. For the convenience in visualization, we multiply the y-axis of the bottom panel by a factor $C(z) \propto H(z)\; (1+z)$ to cancel out redshift factors that carry no extra information once a cosmology is assumed. In this case, the  $\textrm{d}J_{\nu}/\textrm{d}z$ measurements uniquely determine the emissivity $\epsilon_{\nu}$ over the frequency range accessible at a factor of $1+z$ bluewards of the filter bandpass. One can simply deconvolve $\textrm{d}J_{\nu}/\textrm{d}z$ to get $\epsilon_{\nu}$. In fact, this works for the cosmic emissivity of any spectral shape as long as it is not evolving over cosmic time. The spectral resolution is not limited by the filter bandwidth but by the redshift uncertainty in the $\textrm{d}J_{\nu}/\textrm{d}z$ measurements. For the clustering redshift technique, the redshift uncertainty is limited by the correlation length \citep{2015MNRAS.447.3500R}, about 10 comoving Mpc, which can be translated into a spectral resolution $R = \lambda/\Delta \lambda \approx 1/\Delta z \approx 200$--$500$. This is a gain of two orders of magnitude from that of the typical broadband observations. Another way to appreciate the potential constraining power of our technique is that beyond the correlation length, the $\textrm{d}J_{\nu}/\textrm{d}z$ measurements in different redshift bins are independent. Over an appreciable range of redshift, the clustering amplitude can be sampled by hundreds (the number of correlation lengths along the line of sight) of quasi-independent measurements.

In more realistic cases, the cosmic radiation field can be wavelength and redshift dependent. Furthermore, the clustering amplitude bias term $b_{J}$ can also be wavelength and redshift dependent. To further relate these two terms, it is useful to introduce a more fundamental quantity: a rest-frame photon clustering bias $b=b(\nu, z)$ such that  $\delta(\epsilon_{\nu}) = b\, \delta_m$, where $\delta(\epsilon_{\nu}) = \epsilon_{\nu}(x)/\langle\epsilon_{\nu}\rangle -1$ is the 3D overdensity of the spatially varying EBL emissivity. Folding in the bias factor in Equation~\ref{eq:dJdz_to_epsilon}, we thus have
\begin{eqnarray}
\frac{\textrm{d} J_{\nu_{\rm obs}}}{\textrm{d}z}\,b_{J}(z) &=& \frac{\textrm{c}}{4\pi\,H\, (1+z)}\, \nonumber\\ && \int\, \frac{\textrm{d}\nu_{\rm{obs}}}{\nu_{\rm{obs}}}\, R(\nu_{\rm{obs}})\, b(\nu, z)\,\epsilon_{\nu}(\nu, z)\, e^{-\tau}\;,
\label{eq:dJdz_bJ_to_epsilon}
\end{eqnarray}
where the left-hand side is the actual observable that can be obtained with the clustering redshift technique. The effective intensity bias $b_{J}(z)$ is a weighted photon bias seen in the observer frame given by the combination of Equation~\ref{eq:dJdz_to_epsilon} and \ref{eq:dJdz_bJ_to_epsilon}: 
\begin{equation}
b_{J}(z) =  \cfrac{ \displaystyle\int \textrm{d}\nu_{\rm{obs}}\,\nu_{\rm{obs}}^{-1}\, R(\nu_{\rm{obs}})\, b(\nu, z)\,\epsilon_{\nu}(\nu, z)\, e^{-\tau}}{\displaystyle\int \textrm{d}\nu_{\rm{obs}}\,\nu_{\rm{obs}}^{-1}\, R(\nu_{\rm{obs}})\, \epsilon_{\nu}(\nu, z)\, e^{-\tau}}\;.
\label{eq:effective_bJ}
\end{equation}
Equation~\ref{eq:dJdz_bJ_to_epsilon} is key in our generative EBL modeling: for any given EBL emissivity and photon bias on the right-hand side, one can calculate the clustering redshift observable on the left-hand side.
As mentioned above, the product $(\textrm{d}J_{\nu}/\textrm{d}z)\;b_{J}(z)$ can potentially be sampled by hundreds of data points. 
If the number of degrees of freedom of the EBL frequency and redshift dependencies is sufficiently small and if observations in multiple broadbands are available, it is possible to break (some of) the degeneracy between the two terms of the product and constrain separately the emissivity and the clustering bias of the cosmic radiation field.

\subsubsection{Application to the UVB}\label{subsubsec:spectral_tagging_Galex}

\begin{figure}[tp]
    \begin{center}
         \includegraphics[width=0.475\textwidth]{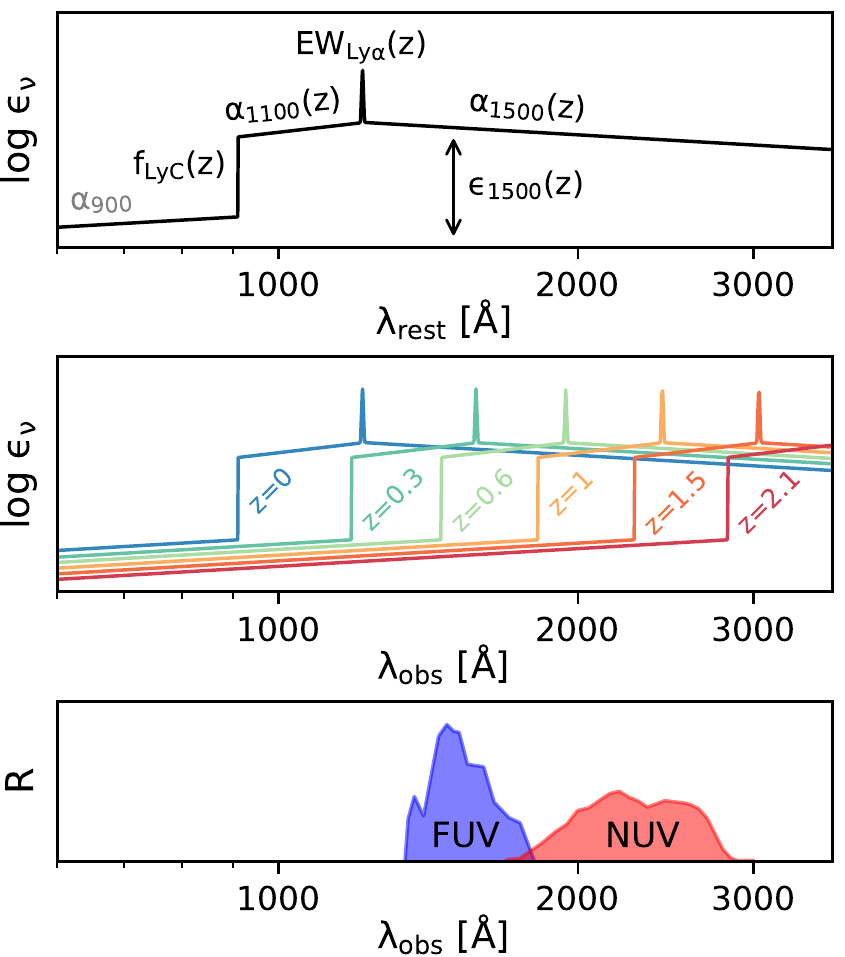}
    \end{center}
    \vspace{-0.75mm}
    \caption{\textbf{Top:} the parameterization of the rest-frame UVB emissivity consists of three segments of the power-law continuum ($\epsilon_{\nu} \propto \nu^{\alpha}$)  with a Lyman break and a Ly$\alpha$ line. We fix the slope $\alpha_{900}$ because the cosmic Lyman continuum is not detected in GALEX. The four spectral features plus one normalization $\epsilon_{1500}$ at $1500\,\AA$ is each allowed to evolve with redshift with one additional parameter, amounting to a total of 10 free parameters. \textbf{Middle:} the emissivity as a function of observer-frame wavelength to show the spectral sampling available for an observer at $z=0$. \textbf{Bottom:} normalized filter response for the FUV and NUV bands of GALEX.}
    \label{fig:CUB_illustration}
\end{figure}

\begin{figure*}[t!]
    \centering
    \includegraphics[width=1\textwidth]{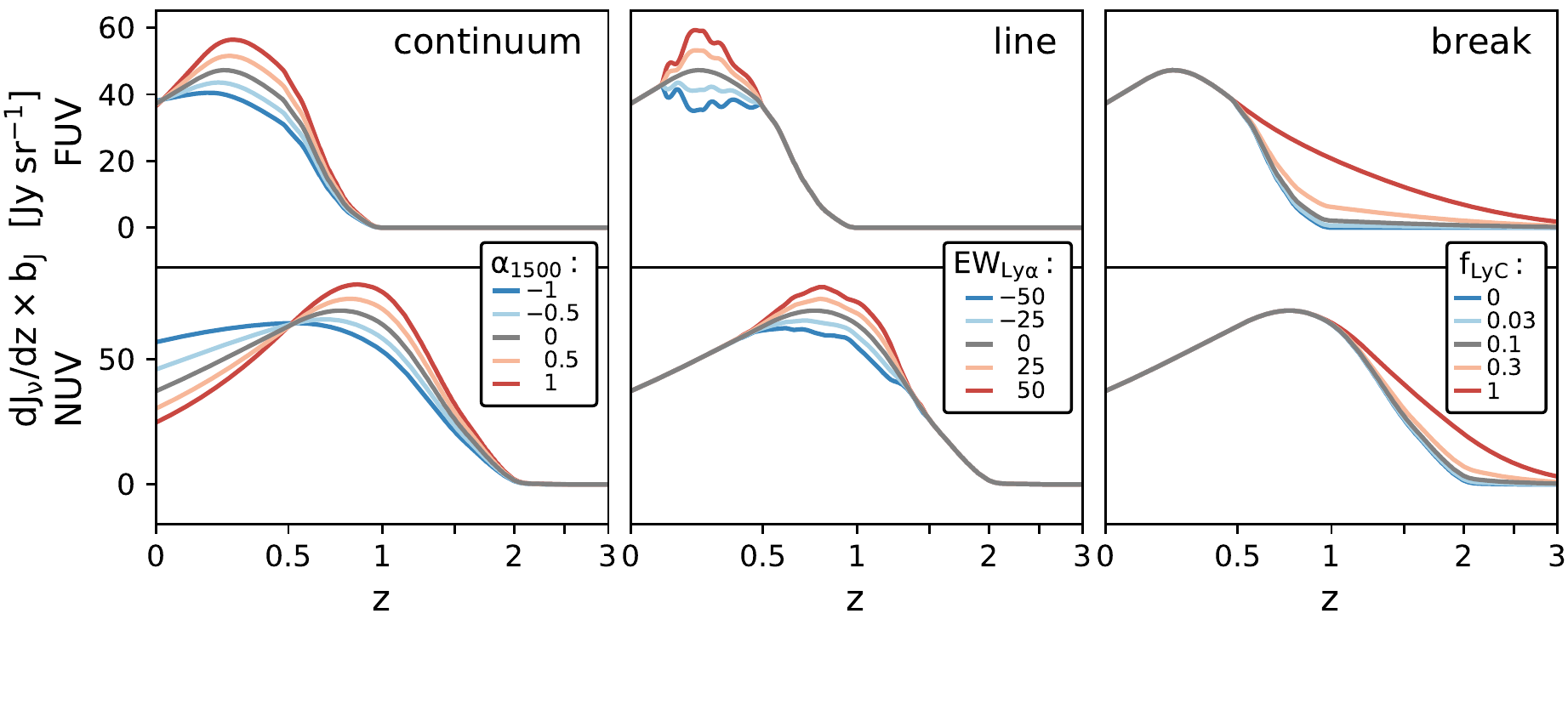}
    \vspace{-9.5mm}
    \caption{Demonstration of tagging spectral features in the UVB emissivity in GALEX broadband tomography. The response in the redshift-deprojected FUV/NUV (top/bottom row) intensities are shown after varying the $1500\,\AA$ continuum slope, Ly$\alpha$ equivalent width, and ionizing photon escape fraction in the left, central, and right columns, respectively. Other parameters are fixed to a set of fiducial values. One can see that these three spectral features trigger different modes of redshift response.}
    \label{fig:model_grid}
\end{figure*}

In this paper, we aim to probe the radiation background over near- to extreme-UV (EUV) at $0<z<2$ using the GALEX All Sky and Medium Imaging Surveys. As a physical spectrum typically shows a high correlation between independent resolution elements, we can reduce the complexity via simple parameterization. We parameterize the volume emissivity with a piecewise power-law function with a Ly$\alpha$ line and Lyman break as shown in the top panel of Figure~\ref{fig:CUB_illustration}. This includes spectral slopes $\alpha$ (where $\epsilon_{\nu} \propto \nu^{\alpha}$) at $900$, $1100$, and $1500\,\AA$, with the first one fixed as we do not have enough signal to noise to constrain the faint ionizing continuum. These two non-ionizing continuum slopes, together with the Ly$\alpha$ equivalent width $\textrm{EW}_{\rm Ly\alpha}$, the strength of the Lyman break or the Lyman continuum escape fraction $f_{\rm LyC}$ and the normalization $\epsilon_{1500}$ at $1500\AA$ are each allowed to evolve with redshift with one additional parameter (see Figure~\ref{fig:parameters_vs_z}). These summed up to a total of 10 free parameters, representing a minimum description of the metagalactic UVB spectrum motivated by atomic physics but without assuming the nature of the source populations (galaxies, quasars, mass-to-light relations, etc.). For simplicity, we have ignored other emission and/or absorption lines like OVI and CIV, which could potentially be present but are likely much fainter compared to Ly$\alpha$ and the continuum for both emissions from galaxies \citep{2018ApJ...863...14B} and the IGM \citep{2013MNRAS.430.3292B}. The equations fully describing our UVB parameterization are given in Appendix~\ref{App:parameterization}.

With the two broadbands onboard of GALEX---FUV ($1350$--$1750\; \AA$) and NUV ($1750$--$2800\;\AA$) shown in the bottom panel of Figure~\ref{fig:CUB_illustration}, we can continuously sample the UVB at different rest-frame frequencies. The accessible spectral sampling starts from the non-ionizing continuum at $z=0$ to the ionizing continuum at $z=1$ in FUV and $z=2$ in NUV as illustrated in the middle panel of Figure~\ref{fig:CUB_illustration}. 

We also parameterize the unknown bias factor in Equation~\ref{eq:dJdz_bJ_to_epsilon} and will fit it simultaneously with the emissivity using the redshift tomographic intensity measurements. We consider a simple 2D power law with three free parameters for the bias factor
\begin{eqnarray}
b(\nu, z) = b_{1500}^{z=0}\,\Bigg(\cfrac{\nu}{\nu_{1500}}\Bigg)^{\gamma_{b\nu}}\, (1+z)^{\gamma_{bz}}\;, 
\label{eq:par_as_fn_of_z_d}
\end{eqnarray}
where $b_{1500}^{z=0}$ is the normalization at $z=0$ at $1500\;\AA$, and $\gamma_{b\nu}$ and $\gamma_{bz}$ are the power indices of its frequency and redshift dependence, which are assumed to be separable. 

Using solely the $(\textrm{d}J_{\nu}/\textrm{d}z)\;b_{J}(z)$ measurements, there is a complete degeneracy between the emissivity and bias normalizations $\epsilon_{1500}^{z=0}$ and $b_{1500}^{z=0}$, such that only their product can be constrained (see Equation~\ref{eq:dJdz_bJ_to_epsilon}). We will break this degeneracy later by using an additional observational constraint from the total intensity in detected sources. For our redshifted tomographic model of the UVB, we effectively have a total of 12 free parameters (10 in the emissivity plus three in the bias minus one normalization degeneracy).

To gain intuition on the spectral tagging using GALEX data, Figure~\ref{fig:model_grid} shows a grid of models in the observable space. The redshift-deprojected FUV (top row) and NUV (bottom row) intensities are shown after varying the non-ionizing UV slope $\alpha_{1500}$, Ly$\alpha$ equivalent width $\textrm{EW}_{\rm Ly\alpha}$, and ionizing photon escape fraction $f_{\rm LyC}$ of the background emissivity one at a time in the left, central, and right columns, respectively. We fix the other parameters to a set of fiducial values: [$\rm{log}\,(\epsilon_{1500}^{z=0}\,b_{1500}^{z=0})$, $\gamma_{\epsilon 1500}$, $\alpha_{1500}^{z=0}$, $C_{\alpha 1500}$, $\alpha_{1100}^{z=0}$, $C_{\alpha 1100}$, $\textrm{EW}_{\rm Ly\alpha}^{z=0.3}$, $\textrm{EW}_{\rm Ly\alpha}^{z=1}$, $\rm{log}\,f_{\rm LyC}^{z=1}$, $\rm{log}\,f_{\rm LyC}^{z=2}$, $\gamma_{b\nu}$, $\gamma_{bz}$] $=$ [25, 2, 0, 0, $-3$, 0, 0, 0, $-5$, $-5$, 0, 1]. Interestingly, these three types of spectral features, i.e., continuum, line, and break trigger different modes of redshift response and at different redshift intervals in these two bands. One can thus unambiguously separate the effects of these main spectral features in the data space. For simplicity, in this example, we set $\alpha_{1500}$, $\textrm{EW}_{\rm Ly\alpha}$, and $f_{\rm LyC}$ constant over cosmic time, but one can see that a joint modeling in two bands does allow us to constrain their first-order redshift evolution as these features are being sampled twice at two different redshift intervals.

\begin{figure*}[!htbp]
    \centering
         \includegraphics[width=0.97\textwidth]{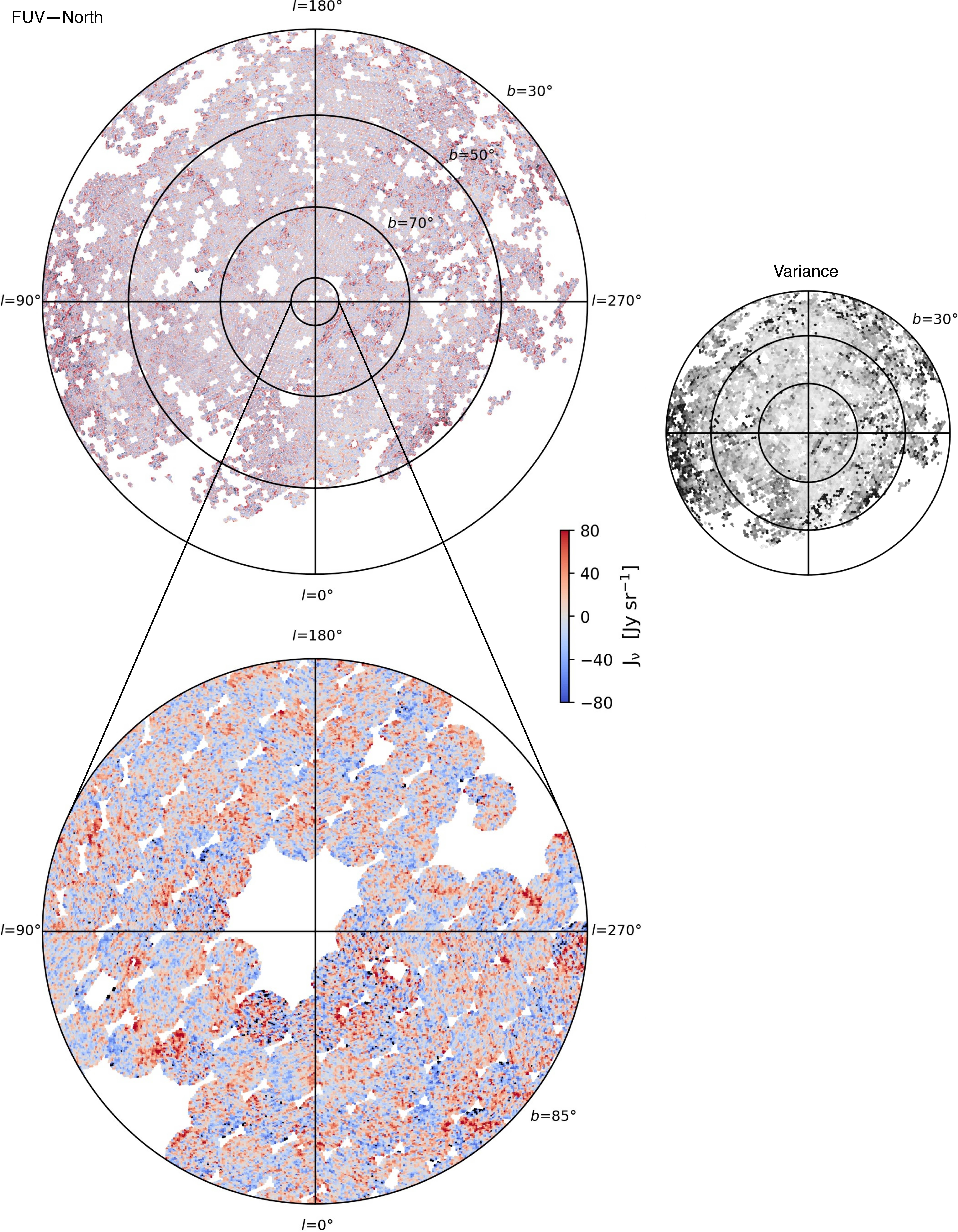}
    \vspace{4mm}
    \caption{GALEX diffuse background anisotropy in FUV displayed using an equal-area Lambert projection. Color maps show the sigma-clipped, tile-median subtracted intensity at $b>30^{\circ}$ (top) and the zoom in near the north Galactic pole at $b>85^{\circ}$ (bottom). Tiles with high foreground with $E(B-V)>0.05$ mag have been removed. Small, gray-scale maps show the per-tile variance to be used for optimal weighting in our cross-correlation measurements. 
    }
    \label{fig:diffuse_maps}
\end{figure*}

\section{Data analysis}\label{sec:data}

\subsection{GALEX intensity maps}\label{subsec:data-Galex}

GALEX is a satellite mission designed to perform wide-field imaging and grism spectroscopy in the UV \citep{2005ApJ...619L...1M,2007ApJS..173..682M}. In this paper, we use its AIS and MIS in the data release GR6/GR7 with observations taken over 2003--2012. The surveys cover a large fraction of the sky above the Galactic plane in two broadbands, FUV ($1350$--$1750\;\AA$) and NUV ($1750$--$2800\;\AA$), down to varying point source depths of AB magnitude 20.5--23.5 with a spatial resolution of $5''$--$10''$. Individual pointings have a circular field of view of $1\fdg2$ in diameter, and the exposure in the two bands are done simultaneously. 

In principle, our cross-correlation tomography takes the total intensity as input and does not require source detection. Practically, we do separate the total intensity into two components, one in detected sources and one in diffuse light below the detection limit. This is for different foreground removal schemes in the data processing, as the foregrounds for sources are stars, while that for the diffuse light are dust scatter light and near-Earth airglow. Another consideration is that because low-redshift EBL is preferentially in detected sources while the high-redshift EBL is mostly in the diffuse component, the former actually acts as a noise-inducing foreground for the latter. To better extract the faint, high-redshift component using angular cross-correlations, we keep the sources and diffuse light separated on the map level. 

Our plan for the map making is the following: to ease the process, we build our diffuse light maps in two GALEX bands based on an existing product of \cite{2014ApJS..213...32M} for which detected sources are already masked; our goal for the source maps is thus to simply recover the light in the exact set of masked sources but keep it in a separate set of maps. The two sets, diffuse and sources, thus sum up to the total intensity field recorded by GALEX. Below we describe our map making in detail.

\subsubsection{Diffuse light}
For the diffuse light, we start from the product generated by \cite{2014ApJS..213...32M}, who masked out all sources detected by the GALEX survey team pipeline, and rebin the images to pixels of $2'$. 
In this product, an attempt to remove the zodiacal light and geocoronal oxygen airglow has been made utilizing the variation seen toward the same patches of the sky as a function of time and location of the spacecraft \citep{2014Ap&SS.349..165M}. Their final diffuse radiation maps in FUV and NUV are dominated by starlight scattered by Milky Way dust, especially at low latitudes. At high latitudes, the extragalactic contribution is significant but its amplitude is under debate, due to the uncertainties in the near-Earth and Galactic foregrounds \citep{2013ApJ...779..180H,2015ApJ...798...14H,2018ApJ...858..101A}. Our cross-correlation analysis has the advantage that the result should not be biased by the presence of foregrounds, as they only add noise but do not correlate with extragalactic large-scale structures. To obtain a random sampling of the sky, we keep only tiles from the AIS or MIS programs and exclude those observed as part of other guest observer programs targeting preselected sources.

We postprocess these data from \cite{2014ApJS..213...32M} with aggressive cleaning for our analysis. We first trim the per-tile field of view from $1\fdg2$ diameter to $1^{\circ}$ to reduce edge effects. For each tile we discard outlier pixels associated with ghosts, bright dust cirrus, or other artifacts by carrying out a 3$\sigma$ clipping in the intensity in FUV and NUV separately. Because the Galactic foreground in the UV correlates strongly with other dust observables, we further remove tiles whose median Galactic reddening (as measured in $E(B-V)$ in \citealt{1998ApJ...500..525S}) is above 0.05 mag. This restricts our analysis to the area with Galactic latitude $|b| \gtrsim 30^{\circ}$. We remove a small number of tiles with highly asymmetric intensity distribution whose mean-to-median ratio or standard deviation to 68th percentile range ratios differ substantially from unity. Finally, because the Galactic foreground fluctuates on all, especially at large scales but, the extragalactic information we wish to extract is primarily at subtile scales, we subtract the median intensity in each tile to get a flat cross-tile zero point. As our cross-correlation estimator (Equation~\ref{eq:w_theta}) uses the absolute but not fractional intensity fluctuations, at this point we do not need to know the amplitude of the monopole extragalactic background being subtracted together with other foregrounds.

We combine all of our selected and processed tiles using the HEALPix scheme \citep{2005ApJ...622..759G}. An $N_{\rm side}$ of 4096 with $50''$ pixels is used to resample the $2'$ pixels in the \cite{2014ApJS..213...32M} images. We show the processed FUV diffuse background map in the northern sky in Figure~\ref{fig:diffuse_maps}. After masking the area outside the footprint of our cross-correlation reference objects (see the next subsection), the final diffuse intensity maps used in our UVB measurement cover about $5500$ deg$^2$ (4500/1000 deg$^2$ in the northern/southern hemisphere) for both FUV and NUV. 

In addition to the intensity maps in two bands, we also construct a set of corresponding error maps to be used as the optimal inverse-variance weighting for our cross-correlation estimator. Assuming large-scale homogeneity for the UVB, the spread of the intensity distribution within each tile thus reflects the level of noise, which is spatially variant due to both varying exposure time and foregrounds. We calculate the per-tile variance based on its 68th percentile range and combine them using the same HEALPix scheme. These error/variance/weight maps evvectively have a resolution of $1^{\circ}$. The FUV variance map over the northern sky is shown together with the intensity in Figure~\ref{fig:diffuse_maps}; one can visually see the correlation between the two.

\subsubsection{Light in detected sources}
\label{subsec:light_in_sources}
The GALEX pipeline provides a catalog of sources combining those detected in the FUV and NUV bands (the ``mcat''). These sources are the ones that have been masked in the \cite{2014ApJS..213...32M} diffuse maps used above. To recover the total light but be able to perform the redshift tomographic cross-correlations separately, we construct a set of intensity maps in FUV and NUV summing up photons in detected sources. Objects brighter than 20 mag but unresolved in GALEX are excluded as they are more likely to be stars, which do not correlate with the extragalactic sky but will add noise to our correlation measurements. For spatial sampling, we treat all objects as point sources and attribute the total flux density of a source to its centroid. The combined flux density field is then placed onto a HEALPix grid of $N_{\rm side}=4096$ and converted into the same intensity unit used in the diffuse maps. We keep the AIS and MIS coverage separated in two maps for each band. Compared to the diffuse light maps, these source maps are spatially sparse especially for the shallower AIS maps where the surface density of sources is low; the source maps are also much less subject to foreground contamination especially at high latitudes. Our source maps differ from the typical galaxy density field used in other large-scale structure studies as ours are light weighted; we expect this to skew the redshift distribution to a more bottom-heavy one by a factor roughly scaled with the luminosity distance. We use the same inverse variance maps built using diffuse light.

\subsection{Sloan Digital Sky Survey (SDSS) large-scale structure reference}\label{subsec:data-SDSS}

Our intensity cross-correlation tomography requires a reference sample of matter tracers in the cosmic web with known redshifts. For this purpose, we combine four spectroscopic samples of galaxies and quasars from the SDSS. At $z\lesssim 0.2$, we take the ``MAIN'' galaxy sample from the NYU value-added catalog made for large-scale structure studies \citep{2005AJ....129.2562B}. Over $0.1\lesssim z \lesssim 0.4$ and $0.4 \lesssim z \lesssim 0.7$, respectively, we rely on the BOSS ``LOWZ'' and ``CMASS'' luminous red galaxy samples. These are from the large-scale structure catalogs built in \cite{2016MNRAS.455.1553R}. At higher redshift, all of our reference objects are from the SDSS DR14 quasar catalog, which is an incremental release containing all SDSS I--III quasars as well as the new objects being obtained by the SDSS IV eBOSS survey. To ensure reliable redshifts for the quasars, we further select those without the $z$-warning flag set. The combined reference sample has a total of about 1.5 million objects within the footprint of the GALEX maps that we build. The reference catalog we use here is very similar to that used in \cite{2019ApJ...870..120C} but includes both the northern and southern SDSS fields. We refer the readers to \cite{2019ApJ...870..120C} for the redshift distribution of each subsample and their redshift-dependent bias factors with respect to matter clustering.

\begin{figure*}[t!]
    \centering
    \includegraphics[width=0.815\textwidth]{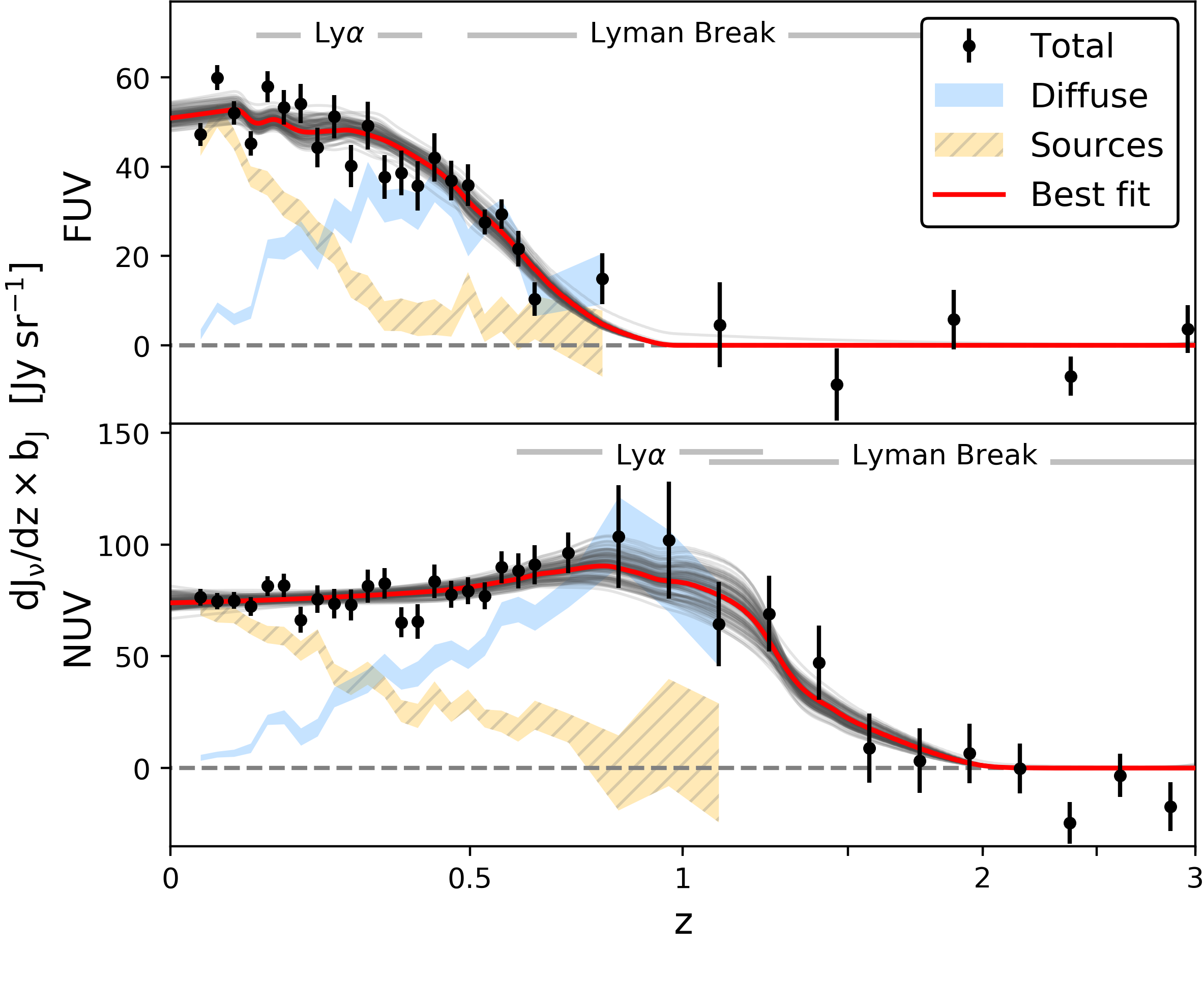}
    \vspace{-4mm}
    \caption{Redshift deprojected, clustering bias-weighted GALEX FUV (top) and NUV (bottom) intensities obtained via the clustering redshift technique. Blue and yellow-hatched bands show the 1$\sigma$ range of the contribution from diffuse light and light in detected sources (down to 20.5--23.5 mag). Data points show the total, which is dominated by diffuse light at high redshifts. The best-fit model in the data space is shown in red curves, and a random subset of 100 MCMC samples is shown in gray curves.}
    \label{fig:dJdz_b}
\end{figure*}

\section{Results}\label{sec:results}

\subsection{Redshift tomography of GALEX maps}\label{sec:dIdz}

Here we present our clustering-based redshift deprojection of the background intensity in the GALEX FUV and NUV bands. For each band and for both the diffuse and detected source maps, we measure $w_{Jr}(\theta, z)$, the angular cross-correlation functions between the intensity field and the reference sample as a function of redshift of the latter as defined in Equation~\ref{eq:w_theta}. To increase the signal-to-noise ratio of our correlation estimator, we use an inverse variance-weighted mean for the ensemble average in Equation~\ref{eq:w_theta}, where the variance is given by the GALEX error maps introduced in Section~\ref{subsec:data-Galex}. We estimate the scale-integrated amplitude $\bar{w}_{Jr}(z)$ by summing up the measured clustering amplitudes over the range $0.5$--$5$ physical Mpc using a power-law angular weighting as described in Equation~\ref{eq:wbar}. Lastly, we correct for the known, redshift-dependent matter clustering amplitudes and the bias factors of the reference sample to get $\bar{w}_{Jr}/(b_r\;\bar{w}_{m}) = (\textrm{d} J_{\nu}/\textrm{d}z)\; b_J$ (Equation~\ref{eq:wbar_to_dJdz}), which is the bias-weighted intensity in the observer bandpass emitted per unit redshift interval as a function of redshift. 

We perform a simple Galactic extinction correction on the normalization of these correlation amplitude measurements but not on the map level. This is to avoid the spatially correlated bias, due to extragalactic imprints in the Galactic dust maps currently available \citep{2019ApJ...870..120C}. Using the \cite{1989ApJ...345..245C} extinction law, \cite{2011Ap&SS.335...51B} calculated the band-averaged $\rm A_{\lambda}/E_{B-V} \approx 8$, which is nearly the same in FUV and NUV due to the presence of the ``$2175\;\AA$ bump'' in the NUV. Over the sky area that we use under our inverse variance-weighting scheme, the effective $E_{B-V}$ is about 20 mmag using \cite{1998ApJ...500..525S} measurements rescaled according to \cite{2011ApJ...737..103S}. The Galactic extinction thus has an amplitude of about 0.15mag, independent of the band and redshift. To correct for dust extinction globally, we therefore scale up our measured ($J_{\nu}/\textrm{d}z)\; b_J$ by $15\%$. We also note that our cross-correlation measurements would not be affected by the potential reference galaxy--Galactic foreground correlation induced by dust extinction. This is demonstrated in \cite{2019ApJ...870..120C} by the stringent limit and null detection in the cross-correlation between SDSS galaxies/quasars and an HI-based reddening map.

Figure~\ref{fig:dJdz_b} shows our estimates of the angular cross-correlations between GALEX specific intensity and the density of reference spectroscopic objects as a function of redshift. As described in Section~\ref{subsec:redshift_tomo}, this quantity corresponds to the product $(\textrm{d} J_{\nu}/\textrm{d}z)\; b_J$. The top/bottom panels show the measurements for FUV/NUV in diffuse light (blue bands; 1$\sigma$ range) and detected source (yellow hatched regions) components. The black data points show the sum of these two components. The error bars are estimated by bootstrapping our reference sample and calculating the dispersion in the estimated cross-correlation amplitudes. A $3\%$ cosmic variance error \citep[calculated based on][]{2008ApJ...676..767T} and $3\%$ zero-point error are added in quadrature to the bootstrapping errors. At $z>1/1.5$ in FUV/NUV, we only include the diffuse component in the total, because the source contribution is consistent with zero at those redshifts and only adds noise. We can observe that the redshift dependence of $(\textrm{d} J_{\nu}/\textrm{d}z)\; b_J$ mainly reflects the emission redwards of the Lyman break being present in a given filter. Our clustering-based redshift measurements are revealing spectroscopic features of the background light (i.e., the cosmic $K$-correction). Below, we present quantitative constraints on the redshift-dependent UVB spectrum.

\subsection{Spectral tagging the UVB}\label{subsec:result_spectral_tagging}

\subsubsection{Bayesian inference and Markov Chain Monte Carlo (MCMC)}\label{subsubsec:MCMC}

Given our redshift tomographic measurements, a generative model describing the response in the observable for any given UVB emissivity and bias (Equation~\ref{eq:dJdz_bJ_to_epsilon}) and our specific parameterization, we now constrain the cosmic time- and frequency-dependent UVB under a Bayesian framework. An MCMC method will be used to obtain the posterior probability distributions of model parameters. Here we describe the ingredients of our inference as follows.

\begin{enumerate}

    \item \textbf{Data \textit{D}}: the primary dataset we use to constrain the model is the redshift deprojected, bias-weighted FUV and NUV intensities $(\textrm{d} J_{\nu}/\textrm{d}z)\; b_J$ shown in Figure~\ref{fig:dJdz_b}. Ideally, one would also include the total, redshift-projected extragalactic monopole intensities as additional integral constraints. However, in the UV, the amplitudes of the monopoles are still under debate. We therefore use only the ratio of the monopoles in NUV versus FUV, which is better known. We use the value $J_{\nu}^{NUV}/J_{\nu}^{FUV}=3\pm 0.3$ based on analyses in the integrated galaxy light (IGL) down to faint magnitudes \citep{2005ApJ...619L..11X,2016ApJ...827..108D}. This effectively appends one data point to our data vector, 
\begin{equation}
 \textbf{\emph{D}}= \Bigg(\frac{\textrm{d} J^{FUV}_{\nu}}{\textrm{d}z}\, b_J(\textbf{\emph{z}}),\ \frac{\textrm{d} J^{NUV}_{\nu}}{\textrm{d}z}\, b_J(\textbf{\emph{z}}),\ \frac{J_{\nu}^{NUV}}{J_{\nu}^{FUV}}\Bigg) \;, 
\end{equation}
where $\textbf{\emph{z}}$ is the redshift bin vector.

    \item \textbf{Model \textit{M}}: as laid out in Section~\ref{sec:method}, our model $\textbf{\emph{M}}$ involves the functional form of $\epsilon_{\nu}(\nu, z)$ (see Appendix~\ref{App:parameterization}) and $b(\nu, z)$ (Equation~\ref{eq:par_as_fn_of_z_d}) and how they relate to the observables in the data space based on radiative transfer (Equation~\ref{eq:J_nu}, \ref{eq:J_nu_Broadband}, and \ref{eq:dJdz_bJ_to_epsilon}). This includes the filter response functions taken from \cite{2005ApJ...619L...7M} and the amount of IGM absorption with the optical depth taken from the analytic approximation in \cite{2014MNRAS.442.1805I} based on observations of intervening neutral clouds seen as absorption line systems in quasar spectra. There are 12 free parameters in our model: 
\begin{eqnarray}
\boldsymbol{\theta} =&& \Big(\rm{log}\,(\epsilon_{1500}^{z=0}\,b_{1500}^{z=0}),\ \gamma_{\epsilon 1500},\ \alpha_{1500}^{z=0},\ C_{\alpha 1500},\nonumber\\&& \alpha_{1100}^{z=0},\ C_{\alpha 1100},\ \textrm{EW}_{\rm Ly\alpha}^{z=0.3},\ \textrm{EW}_{\rm Ly\alpha}^{z=1},\ \rm{log}\,f_{\rm LyC}^{z=1},\ \nonumber\\&& \rm{log}\,f_{\rm LyC}^{z=2},\ \gamma_{b\nu},\ \gamma_{bz}\Big)\;.
\label{eq:theta_pars}
\end{eqnarray}

    \item \textbf{Likelihood P(\textit{D\,|\,}$\boldsymbol{\theta}$\textit{,\,M})}: given a set of parameters $\boldsymbol{\theta}$ that determine the UVB emissivity and clustering bias factor, we calculate the expected data. For $(\textrm{d} J_{\nu}/\textrm{d}z)\; b_J$, we use Equation~\ref{eq:dJdz_bJ_to_epsilon}. For the monopole ratio, we take $J_{\nu}$ in the two bands using Equations~\ref{eq:J_nu} and \ref{eq:J_nu_Broadband}. As our redshift binning is wider than the typical correlation length in redshift space, we treat all the data as independent measurements. Assuming Gaussian errors, the likelihood function is thus
\begin{eqnarray}
  \mathrm{L} &=& \mathrm{P}(\textbf{\emph{D}}\,|\,\boldsymbol{\theta},\ \textbf{\emph{M}})\nonumber\\&=& \prod_i\, \frac{1}{\sqrt{2\pi\,\sigma_i^2}}\, \rm{exp}\,\Bigg(-\frac{(D'_i-D_i)^2}{2\sigma_i^2}\Bigg)\;, 
\end{eqnarray}
where $D'_i$ and $D_i$ are the expected and measured data vectors, respectively, and $\sigma_i$ are the errors in the measurements.

    \item \textbf{Prior P($\boldsymbol{\theta}$)}: we employ flat priors for most of the parameters with a few exceptions. For $C_{\alpha 1500}$ and $C_{\alpha 1100}$ parameterizing the redshift evolution of the corresponding spectral slopes (see Appendix~\ref{App:parameterization}), we do not have strong constraints from our data. We therefore set a wide Gaussian prior of $0\pm1.5$ for each. The redshift evolution power index $\gamma_{\epsilon 1500}$ for the $1500\AA$ emissivity normalization is highly degenerate with spectral slopes $\alpha_{1100}$ and $\alpha_{1500}$, and also $b(\nu, z)$, which can be appreciable in the left panel of Figure~\ref{fig:model_grid} as all these parameters affect the long-range tilt of $(\textrm{d} J_{\nu}/\textrm{d}z)\; b_J(z)$. We expect a strongly rising $1500\AA$ emissivity from $z=0$ to $z=2$ following the cosmic star-formation, or similarly the black hole accretion history \citep{2014ARA&A..52..415M}. Based on direct rest-FUV measurements in detected sources down to faint magnitudes uncorrected for interstellar medium (ISM) dust attenuation \citep{2005ApJ...619L..47S,2016ApJ...832...56A}, we set a Gaussian prior of $2\pm0.3$ in $\gamma_{\epsilon 1500}$. We note that this prior is not model dependent and originates directly from observations. The priors and the ranges allowed for our parameters are summarized in Table~\ref{table:priors}.

\end{enumerate}

\makeatletter
\newcommand{\thickhline}{%
    \noalign {\ifnum 0=`}\fi \hrule height 0.8pt
    \futurelet \reserved@a \@xhline
}
\newcolumntype{"}{@{\hskip\tabcolsep\vrule width 0.8pt\hskip\tabcolsep}}
\makeatother

\begin{table}[t]
\caption{\label{table:priors} \normalsize Priors and posteriors of the parameters}
\begin{center}
\renewcommand{\arraystretch}{1.7}
\begin{tabular}{c"ccc}
\thickhline
Parameter & \multicolumn{2}{c}{Range\,/\,Prior} & Posterior \\ \thickhline
$\rm{log}\,(\epsilon_{1500}^{z=0}\;b_{1500}^{z=0})$ & [20, 30] & flat & $25.13^{+0.01}_{-0.01}$ \\ 
$\gamma_{\epsilon 1500}$ & [-7, 7] & Gaussian $2\pm0.3$ & $2.06^{+0.31}_{-0.30}$ \\
$\alpha_{1500}^{z=0}$ & [-7, 7] & flat  & $-0.08^{+1.28}_{-0.84}$ \\ 
$C_{\alpha 1500}$ & [-7, 7] & Gaussian $0\pm1.5$ & $1.85^{+1.22}_{-1.28}$ \\ 
$\alpha_{1100}^{z=0}$ & [-7, 7] & flat  & $-3.71^{+1.34}_{-0.98}$ \\ 
$C_{\alpha 1100}$ & [-7, 7] & Gaussian $0\pm1.5$ & $0.50^{+1.46}_{-1.44}$ \\ 
$\textrm{EW}_{\rm Ly\alpha}^{z=0.3}$ & [-500, 500] & flat  & $-6.17^{+12.63}_{-11.43}$ \\ 
$\textrm{EW}_{\rm Ly\alpha}^{z=1}$ & [-500, 500] & flat  & $88.02^{+51.44}_{-48.87}$ \\ 
$\rm{log}\,f_{\rm LyC}^{z=1}$ & [-20, 0] & flat & $<-0.53\;(3\sigma)$ \\ 
$\rm{log}\,f_{\rm LyC}^{z=2}$ & [-20, 0] & flat  & $<-0.84\;(3\sigma)$ \\ 
$\gamma_{b\nu}$ & [-7, 7] & flat  & $-0.86^{+0.83}_{-1.29}$ \\ 
$\gamma_{bz}$ & [-7, 7] & flat  & $0.79^{+0.32}_{-0.33}$ \\ \thickhline
\end{tabular}
\end{center}
\end{table}

\begin{figure*}[ht!]
\begin{minipage}[b]{0.483\linewidth}
\centering
\includegraphics[width=\textwidth]{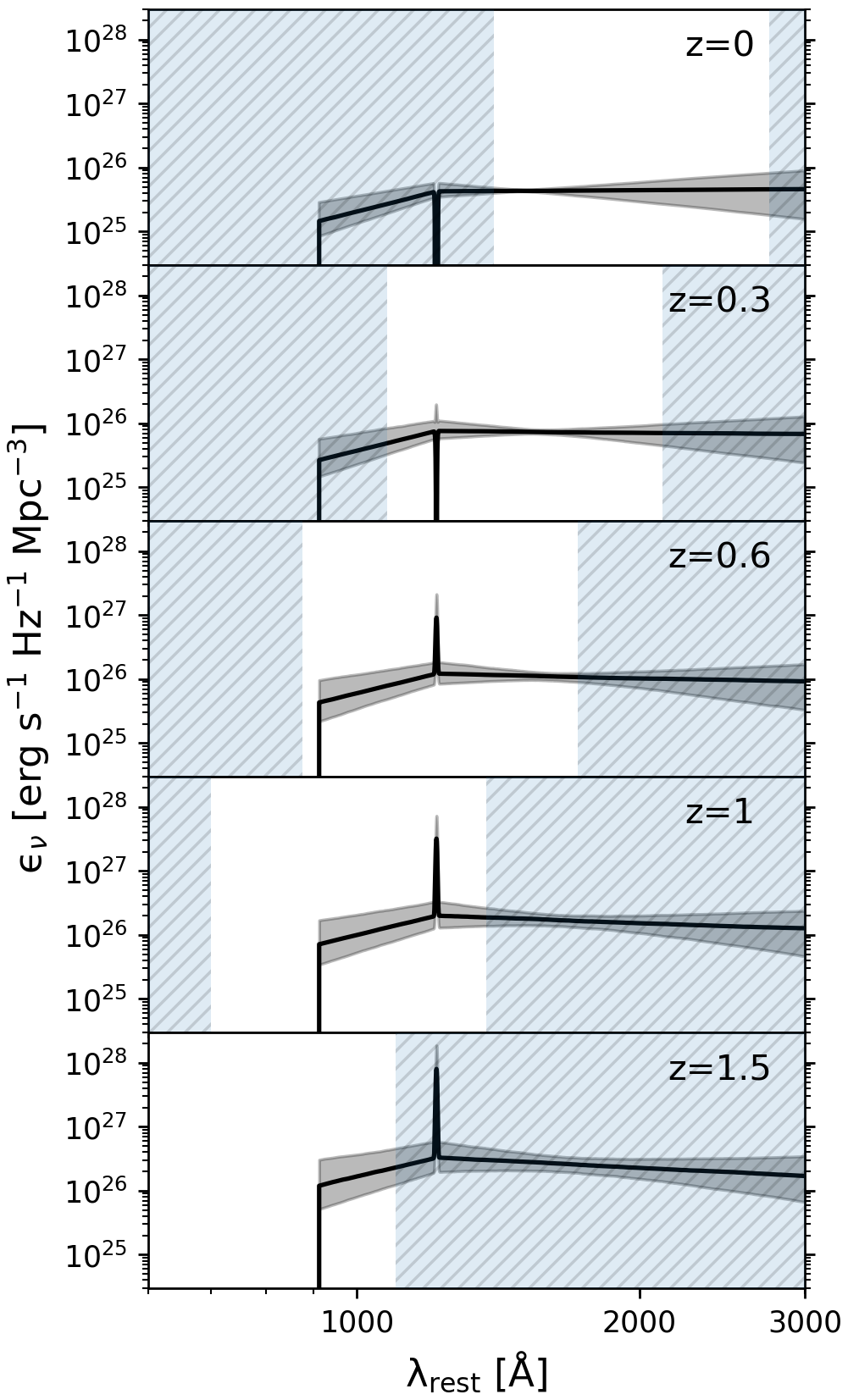}
\caption{Posterior UVB comoving volume emissivity as a function of wavelength and redshift in our GALEX spectral tagging analysis. Black lines show the posterior medians, and gray bands show the $1\,\sigma$ errors. The hatched area indicates regions with no direct data constraint in GALEX; results in these regions should be viewed as extrapolations.}
\label{fig:volume_emissivity}
\end{minipage}
\hspace{0.7cm}
\begin{minipage}[b]{0.474\linewidth}
\centering
\includegraphics[width=\textwidth]{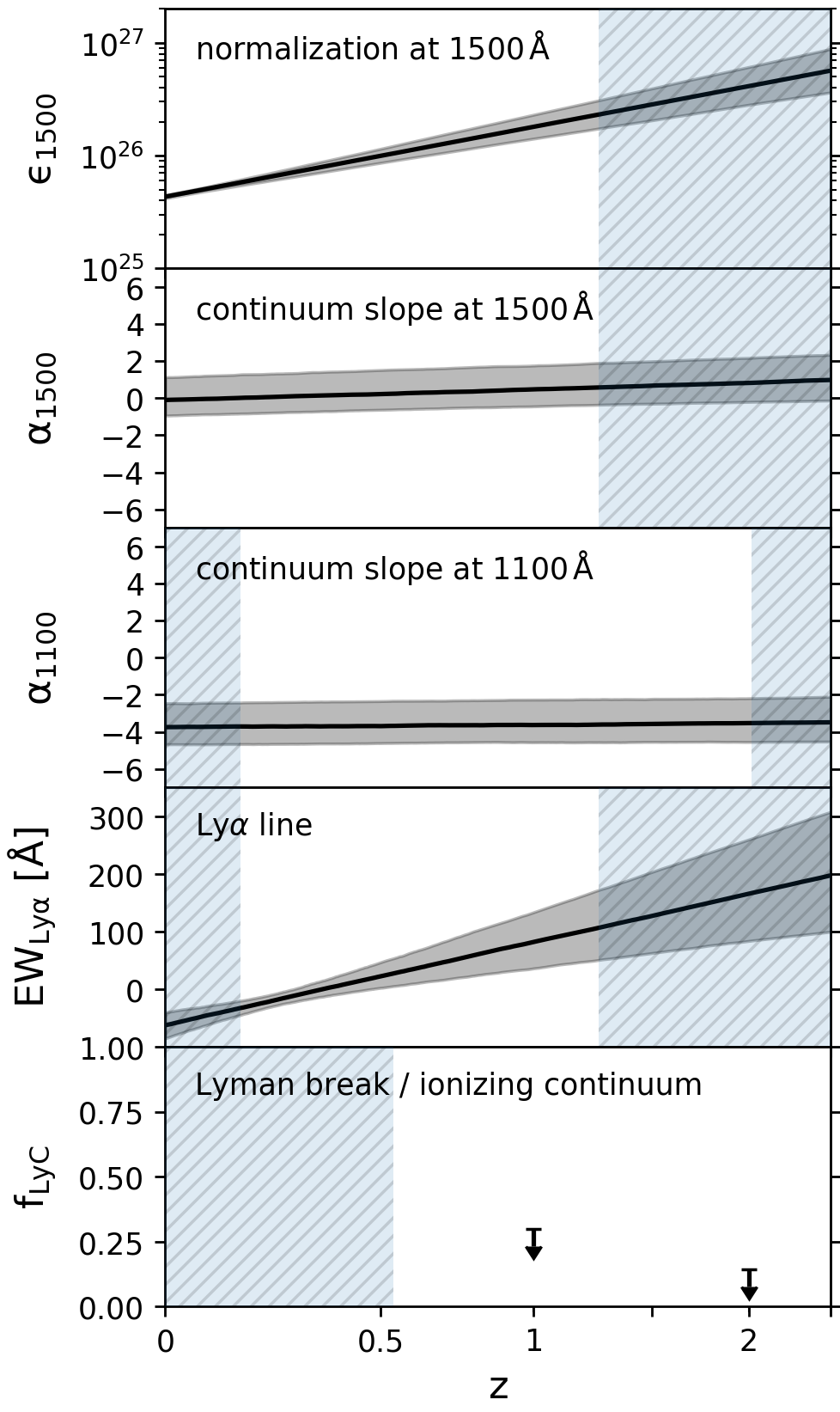}
\caption{Posterior UVB emissivity parameters as a function of redshift. Black lines and gray bands show the posterior median and $1\,\sigma$ errors. Downward arrows show $3\,\sigma$ upper limits. The hatched area indicates regions with no direct data constraint. We note that our linear parameterization in $\textrm{EW}_{\rm Ly\alpha}$ is needed for the positive detection to be robust.
}
\label{fig:parameters_vs_z}
\end{minipage}
\end{figure*}

Having specified all the ingredients in the Bayes' rule, 
\begin{equation}
  \mathrm{P}(\boldsymbol{\theta}\,|\textbf{\emph{D}}) \propto \mathrm{P}(\textbf{\emph{D}}\,|\,\boldsymbol{\theta})\, \mathrm{P}(\boldsymbol{\theta})\;, 
\end{equation}
we use an MCMC package of \cite{2013PASP..125..306F} to sample the posterior distributions $\mathrm{P}(\boldsymbol{\theta}\,|\textbf{\emph{D}})$ of our model parameters given the data. The fitted posterior median and 16th/84th percentiles for each parameter are summarized in Table~\ref{table:priors}. A certain level of covariances between the parameters is present, which we visualize in Appendix~\ref{App:Covariance}.

Figure~\ref{fig:dJdz_b} overlays the measurements with the best-fit (posterior median) model in the data space with red curves. A random subset of 100 MCMC samples is also shown in gray curves. One can see that, overall, the model can sufficiently describe the data. At high redshift, the drop-offs in both bands are due to the presence of the Lyman break. At low redshift before the break comes in, the overall flat or increasing $(\textrm{d} J_{\nu}/\textrm{d}z)\; b_J$ suggests an increasing emissivity and/or clustering bias toward high redshift; otherwise, the steep cosmic dimming factor (i.e. $1/[H(z)\,(1+z)]$ in Equation~\ref{eq:dJdz_bJ_to_epsilon}) would quickly suppress the correlation amplitudes. There is a hint of cosmic Ly$\alpha$ emission present at $z\approx1$, resulting in a small bump when the line is redshifting through the NUV filter. This will be discussed in detail later. Interestingly, one can see that the model tries to match the wiggles in the data at $z\approx0.3$ in FUV. This is possible because we include a sharp feature, i.e., Ly$\alpha$ that is convolved with the filter response at these redshifts (see the middle column in Figure~\ref{fig:model_grid}). This drives the best-fit $\textrm{EW}_{\rm Ly\alpha}^{z=0.3}$ value away from exactly (but still consistent with) zero. The exact shapes of the filter curves are only known to $5$--$10\%$ precision. If improved, we will gain constraining power on the line detection. It is equally interesting that the data in NUV over the same redshift range do not show significant wiggles, and indeed, the model does not allow short-mode fluctuations as the corresponding spectrum is a featureless continuum at $1400$--$2800\AA$; this provides evidence that the wiggles in FUV may be real and not due to the underestimation of the errors.

\subsubsection{Breaking the intensity--bias degeneracy}\label{subsubsec:break_intensity_bias_degeneracy}

We now have gathered enough information to break the intensity--bias degeneracy in the normalization parameter  $\epsilon_{1500}^{z=0}\,b_{1500}^{z=0}$. This utilizes the finding that at $z=0$, almost all of the clustered UVB photons are from detected extragalactic sources (Figure~\ref{fig:dJdz_b}), whose projected monopole intensity $J_{\nu}$ is much better known. Additionally, the contamination from foreground sources (i.e. stars) is low especially at high latitudes. Combining the measured $J_{\nu}$ and  $(\textrm{d}\,J_{\nu}/\textrm{d} z) \, b_{J}$ in detected sources, the relationship between $b_{J}$ and $b(\nu, z)$ given in Equation~\ref{eq:effective_bJ} (which depends on the fitted $\epsilon_{\nu}$), and the integral constraint in Equation~\ref{eq:dJdz_normalization}, one can solve for the bias normalization $b_{1500}^{z=0}$. In our case, we can greatly simplify Equation~\ref{eq:effective_bJ} because no frequency weighting for the filter and emissivity is needed as the fitted spectral slope $\alpha_{1500}^{z=0}$ is about 0 and the IGM absorption at low redshifts for non-ionizing photons is negligible. The effective intensity bias becomes the emitted photon bias evaluated at the observed bands:
\begin{equation}
 b_{J}(z) \approx b(\bar{\nu}, z)\;, 
\end{equation}
where $\bar{\nu}=\bar{\nu}_{\rm{obs}}\,(1+z)$ for $\bar{\nu}_{\rm{obs}}$ as the effective frequency of FUV or NUV. Given our parameterization for $ b(\nu, z)$, we therefore have 
\begin{equation}
 b_{1500}^{z=0} = \cfrac{1}{J_{\nu}^{sources}}\, \displaystyle\int \, \textrm{d} z\,\cfrac{ (\textrm{d}\,J_{\nu}^{sources}/\textrm{d} z)\, b_{J}}{(\bar{\nu}_{\rm{obs}}/\nu_{1500})^{\gamma_{b\nu}}\, (1+z)^{\gamma_{b\nu}+\gamma_{bz}^{sources}}}
\end{equation}
where $J_{\nu}^{sources}$, $= \textrm{d}\,J_{\nu}^{sources}/\textrm{d} z$, $\gamma_{bz}^{sources}$ are, in this case, the values for detected sources instead of those for the total UVB. A flux-limited sample of sources corresponds to an increasingly rarer and highly biased part of the total background emissivity at higher redshift. 
Based on the redshift-dependent luminosity threshold in GALEX and the luminosity-dependent galaxy bias measurements in SDSS from  \cite{2011ApJ...736...59Z}, the bias for a flux-limited source sample scales roughly like $(1+z)^2$, i.e. a factor of $(1+z)$ steeper than that given by our best-fit slope $\gamma_{bz}$ for the total EBL. To properly propagate the covariance with other parameters, we assume that the detected source component follows $\gamma_{bz}^{sources} = \gamma_{bz}+1$; for each MCMC sample of the total UVB posteriors for $\gamma_{bz}$ and $\gamma_{b\nu}$, we can therefore sample the posterior of the bias normalization $b_{1500}^{z=0}$ using the detected source component. The result is $b_{1500}^{z=0} = 0.32 \pm0.05$. Given the rapid decline of the detected source contribution at higher redshift, we note that the precise value of the slope describing the redshift dependence $\gamma_{bz}^{sources}$ has a weak effect on the estimation of the bias parameter $b_{1500}^{z=0}$. Increasing/decreasing the slope $\gamma_{bz}^{sources}$ of the redshift dependence by 0.3 decreases/increases the best-fit $b_{1500}^{z=0}$ by about $5$\% only.

\subsubsection{Cosmic UVB emissivity}\label{subsubsec:result_epsilon_nu}

We now present our results on the UVB volume emissivity $\epsilon_{\nu}(\nu, z)$ as a function of redshift and rest-frame frequency. Figure~\ref{fig:volume_emissivity} shows our posterior $\epsilon_{\nu}(\nu, z)$ at five different redshifts or cosmic time, with the corresponding redshift-dependent spectral parameters shown in Figure~\ref{fig:parameters_vs_z}. In these two figures, the black lines show the posterior median, and the gray bands show the $1\,\sigma$ errors obtained from the 16th/84th percentiles of the MCMC posterior sampling projected onto these one-dimensional spaces. We have obtained the intensity normalization $\epsilon_{1500}^{z=0}$ by dividing the fitted joint normalization $\epsilon_{1500}^{z=0}\,b_{1500}^{z=0}$ by $b_{1500}^{z=0}$ presented in the previous subsection and propagated the error. The hatched area shows regions with no direct data constraint using GALEX bands; the results there are thus considered extrapolations. Out of the 12 parameters (plus the $\epsilon_{1500}^{z=0}\,b_{1500}^{z=0}$ degeneracy now broken), we obtain meaningful constraints on all but three redshift dependency parameters for which external priors are used: $\gamma_{\epsilon 1500}$, $C_{\alpha 1500}$, and $C_{\alpha 1100}$ which describe the relative redshift evolution for the $1500\,\AA$ emissivity, $1500\,\AA$ continuum slope, and $1100\,\AA$ continuum slope, respectively. Our UVB inference is done without any assumption on the nature of the sources involved.

Our analysis constrains some of the key properties of the non-ionizing UVB continuum. These include the amplitude of the overall $1500\,\AA$ emissivity, which quantifies the total radiation output of the universe in this spectral window. The spectral slopes that we obtain at both $1100$ and $1500\,\AA$ probe a combination of the emission and absorption mechanisms in galaxies or quasars before the light enters the metagalactic space into the IGM. For a galaxy-dominated scenario, these spectral slopes are related to the age- and metallicity-dependent stellar populations as well as the absorption in the dusty ISM averaged over all galaxies. We note that although we set a prior of non-evolving $\alpha_{1500}$, the data seem to favor a $\alpha_{1500}$ steepening toward high redshift with marginal significance. The physical implications will be discussed in more detail in Section~\ref{subsec:discussion_Haardt_Madau} together with a widely used synthetic UVB model.

We find clear evidence of the presence of the Lyman break in the UVB, while the leakage of cosmic ionizing photons at $\lambda < 912\, \AA$ is not detected. Our 3$\sigma$ upper limits for the cosmic Lyman continuum escape fraction are placed at $30\%$ and $14\%$ at $z\approx 1$ and 2 probed by the FUV and NUV bands, respectively. The constraint is not particularly tight compared to that obtained using Ly$\alpha$ forest absorption \citep{2003MNRAS.342.1205M,2019MNRAS.486..769K}, but we note that our method is distinct in that it more directly traces the ionizing emission.

It is also interesting to mention the constraints we obtain for the Ly$\alpha$ line. At low redshift, $z\lesssim 0.4$, our estimate of $\textrm{EW}_{\rm Ly\alpha}$ is consistent with zero. At $z\approx1$, we find a $2\sigma$ indication of Ly$\alpha$ emission with $\textrm{EW}_{\rm Ly\alpha} = 88.02^{+51.44}_{-48.87}\; \AA$. This is represented in Figure~\ref{fig:volume_emissivity} using an arbitrary line width for the Ly$\alpha$ set to $\rm FWHM=5\,\AA$ for visualization purposes. Our intensity mapping approach is sensitive to all the Ly$\alpha$ photons from recombination powered by star-formation or black hole accretion, and potential low surface brightness IGM emission. One caveat of our estimate is that if the luminosity-weighted clustering bias factor for Ly$\alpha$ differs from that of the continuum around $1216\, \AA$, the equivalent widths or luminosity densities need to be scaled with the continuum-to-line bias ratio. We will discuss our cosmic Ly$\alpha$ constraints together with that in the literature in Section~\ref{subsec:discussion_Lya}.

\begin{figure}[t]
    \begin{center}
         \includegraphics[width=0.475\textwidth]{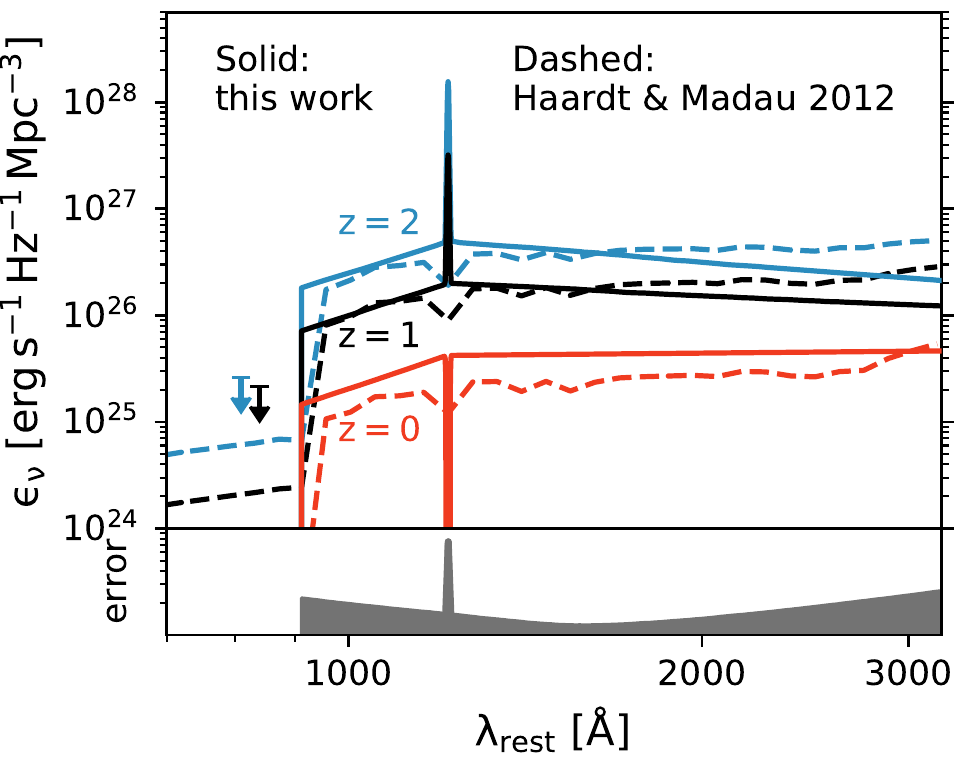}
    \end{center}
    \caption{Comparison between our UVB emissivity measurements with the model of \cite{2012ApJ...746..125H} at $z=0$, 1, and 2. The $1\,\sigma$ error for our measurements (averaged over these three redshifts) is shown in the bottom panel. An overall agreement between the two can be found, while our measurements have a higher $z=0$ normalization, less steep redshift evolution of the normalization, and a hint of hardening in the $1500\,\AA$ continuum slope toward high redshift.}
    \label{fig:compare_to_Haardt}
\end{figure}

\section{Discussion}\label{sec:discussion}

\subsection{Comparison with the Haardt $\&$ Madau model}\label{subsec:discussion_Haardt_Madau}

We compare our direct UVB emissivity measurement to the widely used synthetic model of \citet[][hereafter HM12; see also \citealt{1996ApJ...461...20H,2001cghr.confE..64H}]{2012ApJ...746..125H}. We focus the comparison on the non-ionizing UV continuum, which, in HM12, is dominated by emission from galaxies. HM12 bases its normalization of FUV emissivity on observed luminosity functions. The shape of the UVB spectrum is obtained from a series of modeling including ISM extinction correction \citep{2000ApJ...533..682C}, stellar-population synthesis \citep{2003MNRAS.344.1000B}, cosmic star-formation and metal production history estimations, and transforming back from star formation to a dust-extincted, frequency-dependent emission. Figure~\ref{fig:compare_to_Haardt} shows the UVB emissivity from HM12 in dashed lines compared to our measurements, in solid lines, at $z=0$, 1, and 2. The overall agreement between these two is remarkable. The match in the emissivity amplitudes supports the fidelity of our overall correlation measurements and the clustering bias normalization using the UVB monopole in detected sources. The consistency in the spectral slopes supports both approaches: for our spectral tagging result, it is an empirical measurement with minimum assumptions but has not been applied and tested before; for HM12 the need to invoke stellar population synthesis and dust extinction correction makes their result highly model dependent. As our approach measures the total background, agnostic about the type of sources, the overall agreement with HM12 supports the scenario that the non-ionizing UVB is dominated by galaxies as postulated in HM12.

There are, however, minor differences between the two. Over $0<z<2$ our emissivity evolves with $(1+z)^{2}$, which is shallower than that in HM12 with $(1+z)^{2.6}$. This is driven by different measurements of the FUV luminosity density evolution adopted in HM12 and in the assumed prior of our fitting. Based on the recent compilation of data in \cite{2016ApJ...832...56A} integrated down to a much fainter magnitude limit than previously used, a shallower UV emissivity evolution versus redshift seems to be preferred. 

The HM12 emissivity has almost redshift-invariant spectral slopes, while our measurement shows a mild hardening of the $1500\, \AA$ continuum toward high redshift. This redshift evolution is subtle and is only detected at the $1.5\sigma$ level in our $C_{\alpha 1500}$ parameter, but this is after we set a prior that favors no redshift evolution. At $z=0$, both our and HM12's emissivities have a slope of $\alpha = 0$ in $\epsilon_{\nu}\propto \nu^{\alpha}$ or $\beta=-2$ in $\epsilon_{\lambda}\propto \lambda^{\beta}$ at $1300$--$2800\, \AA$, typical for local starburst galaxies \citep[][]{1999ApJ...521...64M}. At $z=1$, before this spectral range exits our bands, we find a best-fit $\alpha = 0.5$ ($\beta = -2.5$). This bluer UV slope is consistent with the dominant galaxy population being younger, less dusty, and/or less metal enriched \citep{2003MNRAS.344.1000B, 2018ApJ...853...56R} at high redshift. Alternatively, a perhaps counterintuitive scenario is that an increasing contribution from far-IR luminous ``dusty'' galaxies at high redshift could also explain the increasing hardness in the UV; this is because while a substantial amount of light is absorbed, the emerging spectrum is blue and OB-star-dominated if they are not entirely dust enshrouded \citep{2014ApJ...796...95C}. We note that an increasing fractional quasar or active galactic nucleus
(AGN) contribution would not result in a bluer non-ionizing UV continuum at $1300$--$2800\, \AA$ \citep{2001AJ....122..549V}.

\subsection{Cosmic Ly$\alpha$ }\label{subsec:discussion_Lya}

\begin{figure}[t]
    \begin{center}
         \includegraphics[width=0.475\textwidth]{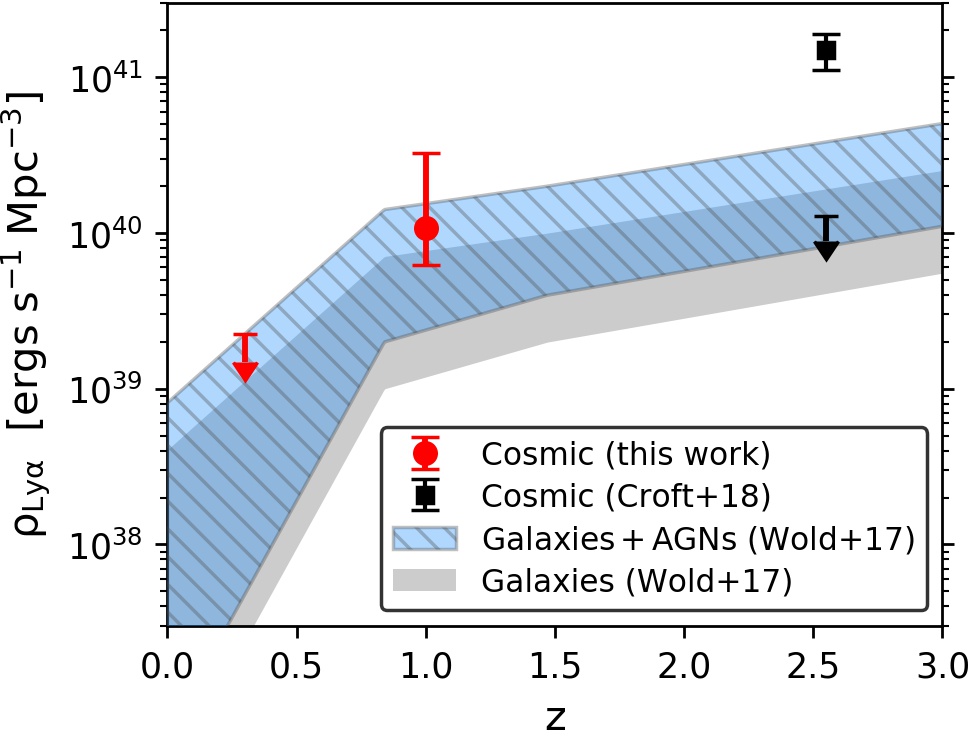}
    \end{center}
    \caption{Ly$\alpha$ luminosity volume density. Our upper limit at $z=0.3$ and detection at $z=1$ are shown by red symbols. The quasar--Ly$\alpha$ and Ly$\alpha$ forest--emission correlation measurements in \citet[][see also \citealt{2016MNRAS.457.3541C}]{2018MNRAS.481.1320C} are shown by the black data point and upper limit, respectively. Gray and blue-hatched bands show the contribution from galaxies and galaxies plus AGNs estimated in \cite{2017ApJ...848..108W}; due to the lack of reliable measurement at $0.4<z<2$, this is obtained by scaling the H$\alpha$ luminosity density measured in the HiZELS survey \citep{2013MNRAS.428.1128S}. At $z=0.3$ and $2<z<3$, this galaxy contribution (gray band) is consistent with direct Ly$\alpha$ emitter survey results \citep{2007ApJ...667...79G,2008ApJS..176..301O,2008ApJ...680.1072D,2010ApJ...711..928C,2010ApJ...714..255G,2011ApJ...736...31B,2012ApJ...744..110C,2016ApJ...823...20K, 2017ApJ...848..108W}.
    }
    \label{fig:Lya_luminosity_density}
\end{figure}

Galaxies and AGNs are known to produce Ly$\alpha$ emission from the recombination of ionized nebula powered by star-formation or supermassive black holes. In addition, fluorescent Ly$\alpha$ powered by ionizing UVB in the diffuse IGM \citep{2010ApJ...708.1048K} or gravitational cooling in the denser part of the collapsing IGM/CGM \citep{2010ApJ...725..633F} might also contribute to the cosmic Ly$\alpha$ budget. In Figure~\ref{fig:Lya_luminosity_density}, we compare our Ly$\alpha$ luminosity density measurements at $z=0.3$ and $z=1$ (red limit/data point) and other results in the literature. Gray and blue hatched bands show the contribution from star-forming galaxies, and galaxies plus AGNs estimated in \cite{2017ApJ...848..108W}. This is obtained via a scaling of the H$\alpha$ luminosity density measured in the HiZELS survey \citep{2013MNRAS.428.1128S}s due to the lack of reliable Ly$\alpha$ luminosity function measurements between $z=0.4$ and $2$. In particular, the current Ly$\alpha$ emitter census at $z\approx1$ using GALEX grism data in NUV is limited to only the brightest sources \citep{2014ApJ...783..119W}. Interestingly, our spectral tagging measurement is consistent with the allowed region for galaxies plus AGN contribution. Because our technique uses no surface brightness thresholding, it is sensitive to potential IGM emission. Our results therefore indicate that the amount of IGM emission cannot be much greater than the total contribution from galaxies and AGNs, i.e., 
\begin{equation}
\rho_{\rm Ly\alpha}^{\rm IGM} / \rho_{\rm Ly\alpha}^{\rm Galaxies+AGN} \lesssim 1\,, 
\end{equation}
valid for both $z=1$ and $z=0.3$. A limited number of theoretical works have explored the predictions of cosmic Ly$\alpha$ from the diffuse IGM at both pre- and post-reionization epochs, while these models currently do not agree with each other quantitatively \citep{2014ApJ...786..111P, 2013ApJ...763..132S, 2016MNRAS.455..725C}.

Figure~\ref{fig:Lya_luminosity_density} also shows the spectroscopic line-intensity mapping results at $z=2.55$ from \cite{2018MNRAS.481.1320C}, who update their earlier measurement in \cite{2016MNRAS.457.3541C}. By cross-correlating quasars and Ly$\alpha$ in the spectra of SDSS galaxy with the best-fit galaxy contribution removed, \cite{2018MNRAS.481.1320C} detected a metagalactic Ly$\alpha$ emission an order of magnitude brighter than that expected from galaxies and AGNs (black data point). In addition, they also use the Ly$\alpha$ forest as the large-scale structure tracer to perform the tracer--spectra correlations, resulting in the null detection of Ly$\alpha$ emission (black upper limit). These authors suggest that the Ly$\alpha$ intensity probed by the quasar--Ly$\alpha$ correlation is not representative for the cosmic mean, but instead it is dominated by reprocessed emission enhanced in the quasar vicinity even at the 1.4--20 Mpc scale. Our technique shares some of the characteristics with that used in \cite{2016MNRAS.457.3541C,2018MNRAS.481.1320C}, but at $z=1$, where our reference objects are also quasars, we do not find an order-of-magnitude higher Ly$\alpha$ emission from expected galaxy contribution. We speculate that our approach might be less subject to this quasar proximity bias for two reasons. First, although we do not probe a larger distance span from the quasars in the transverse dimension on the sky, the line-of-sight distance that we probe is much longer, potentially diluting the effect. Second, the quasar proximity bias might be partly absorbed in the clustering bias factor $b(\nu, z)$ that we fit, again reducing its impact in the emissivity estimations. Of course, it is still possible that either or both studies have yet unidentified systematics.

The Ly$\alpha$ emission of galaxies originates from recombination in HII regions and is usually strongly suppressed by a dusty neutral ISM before escaping to intergalactic space. We define an effective Ly$\alpha$ escape fraction such that 
\begin{equation}
 \rho_{\rm Ly\alpha} = f_{\rm esc}\, C\, \rho_{*}\,, 
\end{equation}
where $\rho_{*}$ is the cosmic star formation rate (SFR) density and $C = 1.1\times10^{42}$ $\rm erg\;s^{-1}\;M_{\odot}^{-1}\;yr$ is the SFR-to-Ly$\alpha$ conversion factor using the empirical H$\alpha$ SFR calibration of \cite{1998ARA&A..36..189K} assuming a Case B recombination Ly$\alpha$-to-H$\alpha$ ratio \citep{2006agna.book.....O}\footnote{This value for $C$ is valid only for the \cite{1955ApJ...121..161S} initial mass function (IMF), but the IMF dependence will be canceled out after being multiplied by $\rho_{*}$, and so would not affect the estimation for $f_{\rm esc}$.}. Given the $\rho_{*}$ measurements compiled in \cite{2014ARA&A..52..415M}, we can place constraints on the cosmic Ly$\alpha$ escape fraction of $f_{\rm esc} < 7\%$ (3$\sigma$) at $z=0.3$ and $f_{\rm esc} = 10_{-6}^{+10}\,\%$ at $z=1$ assuming that all of the Ly$\alpha$ photons originate from galaxies. If AGNs contribute to half of the $\rho_{\rm Ly\alpha}$ we detect, the Ly$\alpha$ escape fractions for galaxies would have to be reduced by a factor of 2. This cosmic effective escape fraction is well within the range of individual Ly$\alpha$- or continuum-selected galaxies \citep{2017ApJ...848..108W,2017ApJ...843..133O}. 

Our estimated UVB Ly$\alpha$ equivalent width of $80\pm50\, \AA$ at $z\approx 1$ is very close to that expected for galaxies with a constant star-formation history based on stellar population synthesis modeling \citep{1993ApJ...415..580C}, but is perhaps on the high side of the distribution for observed galaxies \citep{,2015PASA...32...27H}. As reviewed in \cite{,2015PASA...32...27H}, a limited number of observational results have suggested that the equivalent width of Ly$\alpha$-emitting galaxies might indeed reach its peak at $z=1$ and flatten out toward high redshifts.

\subsection{Total UVB}\label{subsec:discussion_total_background}

The origin and demography of the UV photons contributing to the diffuse light seen in GALEX or earlier UV missions has been a matter of debate \citep{1989ApJ...338..677M, 1991ApJ...379..549M, 1991ARA&A..29...59B, 1991ARA&A..29...89H,2013ApJ...779..180H,2015ApJ...798...14H,2018ApJ...858..101A}. The difficulty arises from the existence of a strong and highly spatially varying component of Galactic dust-scattered light. Even in low dust column density regions of the sky, the detected intensity might still be strongly contaminated by a near-Earth foreground from airglow and zodiacal light \citep{2014Ap&SS.349..165M}, as these UVB measurements have all been done with low-Earth experiments, be it balloon-borne, rocket-borne, or space-based. Recently, \cite{2015ApJ...798...14H} and \cite{2018ApJ...858..101A} argued that after taking into account all the known sources of radiation (Galactic dust, extragalactic background, and near-Earth foreground), there still appears to be a ``mysterious foreground,'' a homogeneous diffuse light component that is of unknown origin.

\begin{figure}[t!]
    \begin{center}
         \includegraphics[width=0.475\textwidth]{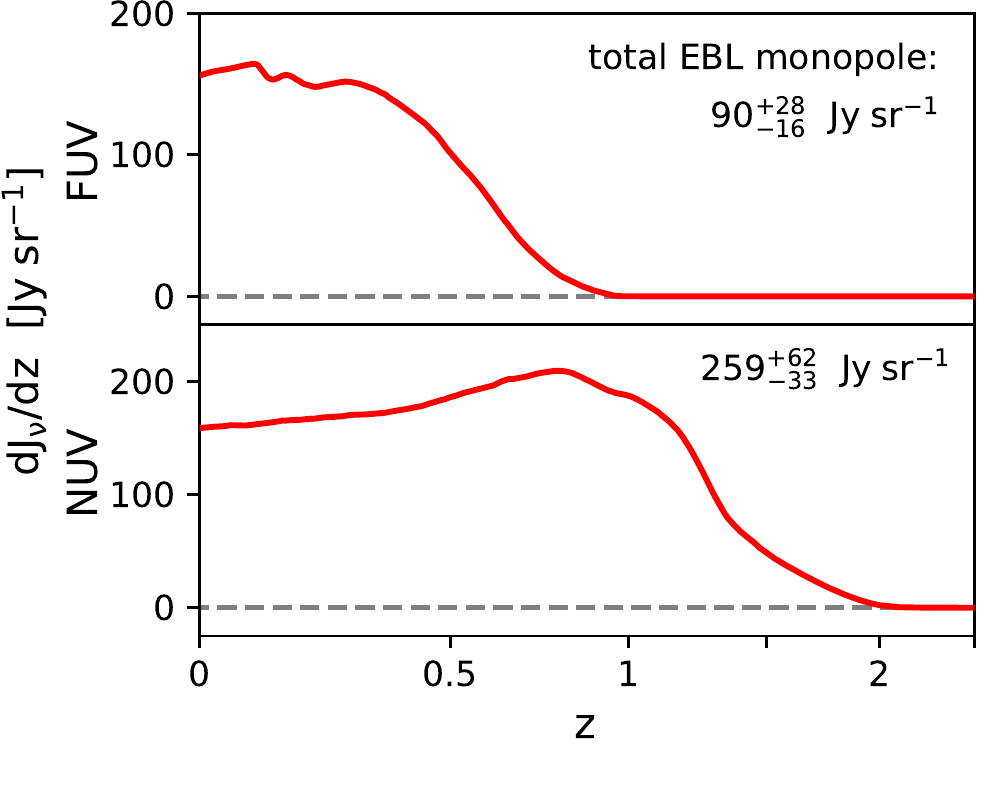}
    \end{center}
    \vspace{-7.5mm}
    \caption{Best-fit UVB intensity in FUV and NUV as a function of emitted redshift. The redshift integral gives the total EBL monopole in each band.}
    \label{fig:dJdz_no_b}
\end{figure}

Our UVB measurement is based on spatial correlations with extragalactic matter tracers. The analysis is insensitive to the presence of foregrounds of non-extragalactic origin, providing a robust constraint on the EBL monopole intensity. In Figure~\ref{fig:dJdz_no_b}, we show our measured $\textrm{d}J_{\nu}/\textrm{d} z$ similar to that in Figure~\ref{fig:dJdz_b} but with the simultaneously fitted bias factor taken out. We integrate the intensity over redshift and find a monopole EBL intensity of $90^{+28}_{-16}$ $\;\rm Jy\;sr^{-1}$ in FUV and $259^{+62}_{-33}$ $\;\rm Jy\;sr^{-1}$ in NUV. These correspond to $89^{+28}_{-16}$ and $172^{+40}_{-21}$ photon units (photons cm$^{-2}$ s$^{-1}$ sr$^{-1}$ $\rm \AA^{-1}$) in FUV and NUV, respectively. We also find that about $30\%$ of the total EBL in both bands is in discrete sources already detected in GALEX AIS and MIS down to $20.5$--$23.5$mag (see Figure~\ref{fig:dJdz_b}). By combining much deeper data from the Hubble Space Telescope with GALEX data, \cite{2016ApJ...827..108D} derived and extrapolated UV luminosity functions to calculate the total IGL (including AGN contribution), resulting in IGL monopoles of $73\pm8$ and $158\pm23$ photon units in FUV and NUV. The differences between our total EBL and the IGL are $16^{+29}_{-18}$ and $14^{+46}_{-31}$ photon units in FUV and NUV, which provide a direct constraint on the cosmic photon production budget allowed for the diffuse IGM. Table~2 summarizes our results on the monopole UVB demographics. 

Summing up our foreground-free estimation of the EBL with the known foregrounds of near-Earth origin \citep{2014Ap&SS.349..165M} and Galactic dust \citep{2018ApJ...858..101A}, there is still a shortage of $200$--$450$ photon units before we can fully explain the total intensity monopole seen in GALEX. Our result thus confirms the existence of an unidentified UV foreground. The nature of this ``mysterious foreground'' remains unknown, but its extragalactic origin can now be firmly ruled out by our clustering analysis.

\newcolumntype{"}{@{\hskip\tabcolsep\vrule width 0.8pt\hskip\tabcolsep}}
\makeatother

\begin{table}[t!]
\normalsize
\caption{\label{tab:total_background} \normalsize Demographics of the UVB monopole}
\begin{center}
\renewcommand{\arraystretch}{1.7}
\begin{tabular}{lcc}
\thickhline
                                     & FUV               & NUV                \\
\multicolumn{1}{l}{}                 & \multicolumn{2}{c}{{[}photon units{]}} \\
\thickhline
Total extragalactic background$^{\rm a}$& $89^{+28}_{-16}$  & $172^{+40}_{-21}$  \\
Galaxies$\,+\,$AGNs, extrapolated$^{\rm b}$         & $73\pm8$          & $158\pm23$         \\
Remaining IGM emission budget        & $16^{+29}_{-18}$  & $14^{+46}_{-31}$   \\
\thickhline
\end{tabular}\\
\end{center}
\vspace{0mm}
\begin{footnotesize}
\-\hspace{0.4cm}$^{\rm a}$ GALEX AIS/MIS, this work\\
\-\hspace{0.4cm}$^{\rm b}$ From \cite{2016ApJ...827..108D}\\
\end{footnotesize}
\end{table}

\subsection{Photon bias and cosmic mass-to-light relation}\label{subsec:discussion_bias}

\begin{figure}[t!]
    \begin{center}
         \includegraphics[width=0.475\textwidth]{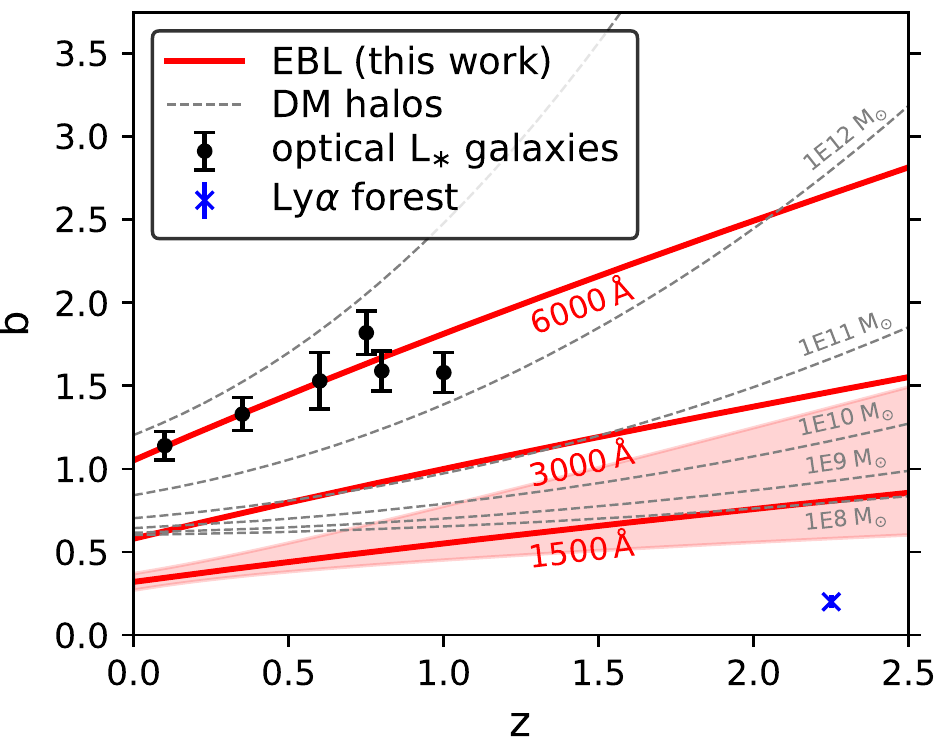}
    \end{center}
    \caption{Linear clustering bias of the background photons in the UV measured in this work and extrapolated into the optical (red lines); the fractional error of about $20\%$ (red band). Gray dashed lines show the halo bias from \cite{2010ApJ...724..878T}. Black data points show the bias for optical $L_{\ast}$ galaxies \citep{2011ApJ...736...59Z, 2013A&A...557A..17M, 2014ApJ...784..128S}. The blue data point shows the bias for the Ly$\alpha$ forest from \cite{2011JCAP...09..001S}.}
    \label{fig:bias_comparison}
\end{figure}

The UVB clustering bias $b(\nu, z)$ contains valuable information about the relation between the sources of radiation and the matter density field, which could be an important summary statistic to constrain cosmological galaxy formation. Based on our definition, $b(\nu, z)$ should be interpreted as the mean clustering bias for photons of rest-frame frequency $\nu$ emitted at redshift $z$. In Figure~\ref{fig:bias_comparison}, we plot our best-fit EBL photon bias
\begin{eqnarray}
b = 0.32\;\Bigg( \frac{\lambda}{1500\,\AA}\Bigg)^{0.86 \pm 1}\, (1+z)^{0.79 \pm 0.3}\;, 
\label{eq:bias_best_fit}
\end{eqnarray}
at 1500 and 3000$\,\AA$ (rest-frame), and extrapolated to 6000$\,\AA$ in the optical. Our result favors the scenario where the EBL bias increases significantly toward high redshift, and at a given redshift, the bias is probably chromatic, with red photons clustered more strongly than blue ones. The uncertainty of this constraint is still large, especially in the frequency dependence. We show the full error for $b$(1500$\,\AA$) in the red band, while both $b$(3000$\,\AA$) and $b$(6000$\,\AA$) are only above $b$(1500$\,\AA$) by 1$\sigma$. Nonetheless, using this relation in optical wavelengths does produce bias values compatible to those measured for optically selected $L_{\ast}$ galaxies \citep[black data points:][]{2011ApJ...736...59Z, 2013A&A...557A..17M, 2014ApJ...784..128S} in both the amplitudes and, more meaningfully, the redshift dependence, given its smaller uncertainty in our constraints. A similar redshift trend has also been reported for galaxies in the UV \citep{2007ApJS..173..503H}. However, a strictly UV-selected galaxy sample at $z<2$ is hard to obtain considering that the process of redshift estimation usually involves matching the objects with optical data. The frequency dependence of our best-fit bias is qualitatively consistent with the low clustering bias found for star-forming galaxies compared to that for red, passive ones \citep{2007ApJS..173..494M,2009ApJ...698.1838H,2008ApJ...672..153C,2017ApJ...838...87C}.

One way to interpret the clustering of the cosmic radiation field is to populate the corresponding sources in dark matter halos using halo models \citep{2002PhR...372....1C}. In Figure~\ref{fig:bias_comparison} we plot the halo bias from the $N$-body simulations compiled in \cite{2010ApJ...724..878T} in gray dashed lines from $10^8$ to $10^{13}\rm\; M_{\odot}$. The UV photon bias is low compared to that of dark matter halos; if both estimations are robust, this would suggest that it may be hard to attribute all of the sources of radiation to galaxies in collapsed halos. We might therefore already see the contribution from a radiation field of more extended and diffuse origin. One extreme example of an uncollapsed matter tracer is the Ly$\alpha$ forest from neutral clouds in the IGM, whose clustering bias is constrained to be very low \citep[$0.2\pm0.04$;][]{{2011JCAP...09..001S}}, as indicated by the blue data point in Figure~\ref{fig:bias_comparison}.

An alternative explanation of the apparent low photon clustering in the UV is the significant level of stochasticity in the cosmic mass-to-light relation. It has been demonstrated that massive, optically selected galaxies like those in our reference sample trace the matter density field with a usually negligible degree of stochasticity on quasi-linear scales \citep{2004MNRAS.355..129S,2016MNRAS.460.1310P}. However, if the UV intensity field does not trace the same underlying large-scale structure sampled by our reference objects, our EBL bias factor $b$ effectively absorbs an EBL--reference correlation coefficient $r_{\epsilon r}$ or EBL--matter correlation coefficient $r_{\epsilon}$ (assuming no stochasticity between the reference and matter). This can be appreciable by examining the form of Equation~\ref{eq:wbar_to_dJdz} where the band-averaged EBL intensity bias is defined. In other words, our estimated bias $b$ is in fact $b = b_0\, r_{\epsilon}$ with $b_0$ being the true EBL clustering bias such that $\delta(\epsilon_{\nu}) = b_0\, \delta_m$; here, the correlation coefficient $r_{\epsilon}$ is defined as
\begin{equation}
r_{\epsilon} = \frac{\langle \delta(\epsilon_{\nu}) \cdot \delta_m \rangle}{\sqrt{\langle \delta(\epsilon_{\nu})^2\rangle\, \langle \delta_m^2 \rangle}}\,, 
\end{equation}
which can be equal to or below unity. If the light-weighted UV sources are, on average, stochastic matter tracers, i.e., $r_{\epsilon}<1$, our measured $b$ would be underestimating the true UVB clustering bias. It remains to be studied what physical mechanisms can create such stochasticity, which might provide insights on the topology of the diffuse IGM and/or galaxy--halo connection in the regime of dwarf or low-surface-brightness galaxies.

\section{Summary}\label{sec:summary}
We present a clustering-based framework to statistically recover frequency and redshift information for the EBL in broadband intensity mapping datasets, and apply it to the GALEX All Sky and Medium Imaging Surveys in the UV. By spatially cross-correlating photons in the FUV and NUV bands with spectroscopic objects in SDSS as a function of redshift, we detect the differential intensity of the UV background (UVB) as a function of redshift up to $z\sim2$. These tomographic measurements clearly reveal imprints of the main spectral features of the UVB redshifting in and out of the bands, allowing us to set empirical constraints on several aspects of the evolving UVB spectrum as follows:

\begin{enumerate}[leftmargin=*]
    \item The overall amplitude and spectral shape of the non-ionizing UVB continuum at $912\,\AA < \lambda < 2700\,\AA$ are in good agreement with the \cite{2012ApJ...746..125H} model. Our results, however, do not rely on any assumption regarding the nature of the sources.
    
    \item Cosmic Ly$\alpha$ emission is tentatively detected with $>95\%$ confidence at $z=1$ with a luminosity density consistent with being powered by cosmic star-formation with an effective escape fraction of 10\%.
    
    \item The Lyman break in the UVB is clearly detected, while the leakage of the cosmic ionizing photons is not detected at $z\sim1$--$3$.
\end{enumerate}

We integrate clustered light over redshift to obtain the total UVB monopoles in the FUV and NUV, which are robust against the presence of foregrounds. These monopoles are in slight excess, but still consistent with the integrated galaxy plus AGN light estimated in \cite{2016ApJ...827..108D}, allowing us to set limits on the cosmic emission from the IGM. Our analysis also provides direct constraints on the photon clustering bias factor as a function of frequency and redshift, which characterizes the cosmic mass-to-light relation. Our GALEX tomography delivers a summary statistic of the net radiation output from cosmological galaxy formation, including the contributions from stars, black holes, and radiative processes in the ISM, CGM, and IGM combined.

This work demonstrates that by combining the concept of intensity mapping, the efficiency of broadband surveys, and a multiband clustering redshift tomography, we can probe the rich astrophysical information in the EBL. The technique can be applied to intensity mapping data of any waveband with any bandwidth.

\begin{acknowledgements}
Y.C. and B.M. acknowledge support from NSF grant AST1313302 and NASA grant NNX16AF64G. We thank Google Cloud for computing support.
\end{acknowledgements}

\bibliographystyle{apj}

\begin{thebibliography}{}

\bibitem[Akshaya et al.(2018)]{2018ApJ...858..101A} Akshaya, M.~S., Murthy, J., Ravichandran, S., Henry, R.~C., \& Overduin, J.\ 2018, \apj, 858, 101 

\bibitem[Alavi et al.(2016)]{2016ApJ...832...56A} Alavi, A., Siana, B., Richard, J., et al.\ 2016, \apj, 832, 56 

\bibitem[Bertone et al.(2013)]{2013MNRAS.430.3292B} Bertone, S., Aguirre, A., \& Schaye, J.\ 2013, \mnras, 430, 3292 

\bibitem[Bianchi(2011)]{2011Ap&SS.335...51B} Bianchi, L.\ 2011, \apss, 335, 51 

\bibitem[Blanc et al.(2011)]{2011ApJ...736...31B} Blanc, G.~A., Adams, J.~J., Gebhardt, K., et al.\ 2011, \apj, 736, 31 

\bibitem[Blanton et al.(2005)]{2005AJ....129.2562B} Blanton, M.~R., Schlegel, D.~J., Strauss, M.~A., et al.\ 2005, \aj, 129, 2562 

\bibitem[Bowyer(1991)]{1991ARA&A..29...59B} Bowyer, S.\ 1991, \araa, 29, 59 

\bibitem[Bruzual \& Charlot(2003)]{2003MNRAS.344.1000B} Bruzual, G., \& Charlot, S.\ 2003, \mnras, 344, 1000 

\bibitem[Byler et al.(2018)]{2018ApJ...863...14B} Byler, N., Dalcanton, J.~J., Conroy, C., et al.\ 2018, \apj, 863, 14 

\bibitem[Calzetti et al.(2000)]{2000ApJ...533..682C} Calzetti, D., Armus, L., Bohlin, R.~C., et al.\ 2000, \apj, 533, 682 

\bibitem[Cantalupo et al.(2014)]{2014Natur.506...63C} Cantalupo, S., Arrigoni-Battaia, F., Prochaska, J.~X., Hennawi, J.~F., \& Madau, P.\ 2014, \nat, 506, 63 

\bibitem[Cardelli et al.(1989)]{1989ApJ...345..245C} Cardelli, J.~A., Clayton, G.~C., \& Mathis, J.~S.\ 1989, \apj, 345, 245 

\bibitem[Casey et al.(2014)]{2014ApJ...796...95C} Casey, C.~M., Scoville, N.~Z., Sanders, D.~B., et al.\ 2014, \apj, 796, 95 

\bibitem[Charlot \& Fall(1993)]{1993ApJ...415..580C} Charlot, S., \& Fall, S.~M.\ 1993, \apj, 415, 580 

\bibitem[Chang et al.(2010)]{2010Natur.466..463C} Chang, T.-C., Pen, U.-L., Bandura, K., \& Peterson, J.~B.\ 2010, \nat, 466, 463 

\bibitem[Cheng et al.(2016)]{2016ApJ...832..165C} Cheng, Y.-T., Chang, T.-C., Bock, J., Bradford, C.~M., \& Cooray, A.\ 2016, \apj, 832, 165 

\bibitem[Chiang \& M{\'e}nard(2019)]{2019ApJ...870..120C} Chiang, Y.-K., \& M{\'e}nard, B.\ 2019, \apj, 870, 120 


\bibitem[Ciardullo et al.(2012)]{2012ApJ...744..110C} Ciardullo, R., Gronwall, C., Wolf, C., et al.\ 2012, \apj, 744, 110 

\bibitem[Coil et al.(2008)]{2008ApJ...672..153C} Coil, A.~L., Newman, J.~A., Croton, D., et al.\ 2008, \apj, 672, 153 

\bibitem[Coil et al.(2017)]{2017ApJ...838...87C} Coil, A.~L., Mendez, A.~J., Eisenstein, D.~J., \& Moustakas, J.\ 2017, \apj, 838, 87 

\bibitem[Comaschi \& Ferrara(2016)]{2016MNRAS.455..725C} Comaschi, P., \& Ferrara, A.\ 2016, \mnras, 455, 725 

\bibitem[Cooray \& Sheth(2002)]{2002PhR...372....1C} Cooray, A., \& Sheth, R.\ 2002, \physrep, 372, 1 

\bibitem[Corlies \& Schiminovich(2016)]{2016ApJ...827..148C} Corlies, L., \& Schiminovich, D.\ 2016, \apj, 827, 148 

\bibitem[Cowie et al.(2010)]{2010ApJ...711..928C} Cowie, L.~L., Barger, A.~J., \& Hu, E.~M.\ 2010, \apj, 711, 928 

\bibitem[Croft et al.(2016)]{2016MNRAS.457.3541C} Croft, R.~A.~C., Miralda-Escud{\'e}, J., Zheng, Z., et al.\ 2016, \mnras, 457, 3541 

\bibitem[Croft et al.(2018)]{2018MNRAS.481.1320C} Croft, R.~A.~C., Miralda-Escud{\'e}, J., Zheng, Z., Blomqvist, M., \& Pieri, M.\ 2018, \mnras, 481, 1320 

\bibitem[Davidsen et al.(1974)]{1974Natur.247..513D} Davidsen, A., Bowyer, S., \& Lampton, M.\ 1974, \nat, 247, 513 

\bibitem[Davis et al.(2018)]{2018MNRAS.477.2196D} Davis, C., Rozo, E., Roodman, A., et al.\ 2018, \mnras, 477, 2196 

\bibitem[Deharveng et al.(2008)]{2008ApJ...680.1072D} Deharveng, J.-M., Small, T., Barlow, T.~A., et al.\ 2008, \apj, 680, 1072 

\bibitem[Driver et al.(2016)]{2016ApJ...827..108D} Driver, S.~P., Andrews, S.~K., Davies, L.~J., et al.\ 2016, \apj, 827, 108 

\bibitem[Faucher-Gigu{\`e}re et al.(2010)]{2010ApJ...725..633F} Faucher-Gigu{\`e}re, C.-A., Kere{\v s}, D., Dijkstra, M., Hernquist, L., \& Zaldarriaga, M.\ 2010, \apj, 725, 633 

\bibitem[Foreman-Mackey et al.(2013)]{2013PASP..125..306F} Foreman-Mackey, D., Hogg, D.~W., Lang, D., \& Goodman, J.\ 2013, \pasp, 125, 306 

\bibitem[Foreman-Mackey et al. (2014)]{corner}
Foreman-Mackey, D., Price-Whelan, A., Ryan, G. et al.\ 2014,
10.5281/zenodo.10598, http://dx.doi.org/10.5281/zenodo.10598

\bibitem[Gnedin \& Ostriker(1997)]{1997ApJ...486..581G} Gnedin, N.~Y., \& Ostriker, J.~P.\ 1997, \apj, 486, 581 

\bibitem[G{\'o}rski et al.(2005)]{2005ApJ...622..759G} G{\'o}rski, K.~M., Hivon, E., Banday, A.~J., et al.\ 2005, \apj, 622, 759 

\bibitem[Gronwall et al.(2007)]{2007ApJ...667...79G} Gronwall, C., Ciardullo, R., Hickey, T., et al.\ 2007, \apj, 667, 79 

\bibitem[Guaita et al.(2010)]{2010ApJ...714..255G} Guaita, L., Gawiser, E., Padilla, N., et al.\ 2010, \apj, 714, 255 

\bibitem[Haardt \& Madau(1996)]{1996ApJ...461...20H} Haardt, F., \& Madau, P.\ 1996, \apj, 461, 20 

\bibitem[Haardt \& Madau(2001)]{2001cghr.confE..64H} Haardt, F., \& Madau, P.\ 2001, Clusters of Galaxies and the High Redshift Universe Observed in X-rays, 64 

\bibitem[Haardt \& Madau(2012)]{2012ApJ...746..125H} Haardt, F., \& Madau, P.\ 2012, \apj, 746, 125 

\bibitem[Haiman et al.(1997)]{1997ApJ...476..458H} Haiman, Z., Rees, M.~J., \& Loeb, A.\ 1997, \apj, 476, 458 

\bibitem[Hamden et al.(2013)]{2013ApJ...779..180H} Hamden, E.~T., Schiminovich, D., \& Seibert, M.\ 2013, \apj, 779, 180 

\bibitem[Hayes(2015)]{2015PASA...32...27H} Hayes, M.\ 2015, PASA, 32, e027 

\bibitem[Heinis et al.(2007)]{2007ApJS..173..503H} Heinis, S., Milliard, B., Arnouts, S., et al.\ 2007, \apjs, 173, 503 

\bibitem[Heinis et al.(2009)]{2009ApJ...698.1838H} Heinis, S., Budav{\'a}ri, T., Szalay, A.~S., et al.\ 2009, \apj, 698, 1838 

\bibitem[Henry(1991)]{1991ARA&A..29...89H} Henry, R.~C.\ 1991, \araa, 29, 89 

\bibitem[Henry et al.(2015)]{2015ApJ...798...14H} Henry, R.~C., Murthy, J., Overduin, J., \& Tyler, J.\ 2015, \apj, 798, 14

\bibitem[Hogg et al.(2002)]{2002astro.ph.10394H} Hogg, D.~W., Baldry, I.~K., Blanton, M.~R., \& Eisenstein, D.~J.\ 2002, arXiv:astro-ph/0210394 

\bibitem[Humason et al.(1956)]{1956AJ.....61...97H} Humason, M.~L., Mayall, N.~U., \& Sandage, A.~R.\ 1956, \aj, 61, 97 

\bibitem[Inoue et al.(2014)]{2014MNRAS.442.1805I} Inoue, A.~K., Shimizu, I., Iwata, I., \& Tanaka, M.\ 2014, \mnras, 442, 1805 

\bibitem[Kennicutt(1998)]{1998ARA&A..36..189K} Kennicutt, R.~C., Jr.\ 1998, \araa, 36, 189 

\bibitem[Khaire et al.(2019)]{2019MNRAS.486..769K} Khaire, V., Walther, M., Hennawi, J.~F., et al.\ 2019, \mnras, 486, 769 

\bibitem[Kollmeier et al.(2010)]{2010ApJ...708.1048K} Kollmeier, J.~A., Zheng, Z., Dav{\'e}, R., et al.\ 2010, \apj, 708, 1048 

\bibitem[Konno et al.(2016)]{2016ApJ...823...20K} Konno, A., Ouchi, M., Nakajima, K., et al.\ 2016, \apj, 823, 20 

\bibitem[Kovetz et al.(2017)]{2017arXiv170909066K} Kovetz, E.~D., Viero, M.~P., Lidz, A., et al.\ 2017, arXiv:1709.09066 

\bibitem[Lesgourgues(2011)]{2011arXiv1104.2932L} Lesgourgues, J.\ 2011, arXiv:1104.2932

\bibitem[Limber(1953)]{1953ApJ...117..134L} Limber, D.~N.\ 1953, \apj, 117, 134 

\bibitem[Madau(1992)]{1992ApJ...389L...1M} Madau, P.\ 1992, \apjl, 389, L1 

\bibitem[Madau \& Dickinson(2014)]{2014ARA&A..52..415M} Madau, P., \& Dickinson, M.\ 2014, \araa, 52, 415 

\bibitem[Maller et al.(2005)]{2005ApJ...619..147M} Maller, A.~H., McIntosh, D.~H., Katz, N., \& Weinberg, M.~D.\ 2005, \apj, 619, 147 

\bibitem[Martin \& Bowyer(1989)]{1989ApJ...338..677M} Martin, C., \& Bowyer, S.\ 1989, \apj, 338, 677 

\bibitem[Martin et al.(1991)]{1991ApJ...379..549M} Martin, C., Hurwitz, M., \& Bowyer, S.\ 1991, \apj, 379, 549 

\bibitem[Martin et al.(2005)]{2005ApJ...619L...1M} Martin, D.~C., Fanson, J., Schiminovich, D., et al.\ 2005, \apjl, 619, L1 

\bibitem[Martin et al.(2014)]{2014ApJ...786..107M} Martin, D.~C., Chang, D., Matuszewski, M., et al.\ 2014, \apj, 786, 107 

\bibitem[Marulli et al.(2013)]{2013A&A...557A..17M} Marulli, F., Bolzonella, M., Branchini, E., et al.\ 2013, \aap, 557, A17 

\bibitem[Matthews \& Newman(2010)]{2010ApJ...721..456M} Matthews, D.~J., \& Newman, J.~A.\ 2010, \apj, 721, 456 

\bibitem[McQuinn \& White(2013)]{2013MNRAS.433.2857M} McQuinn, M., \& White, M.\ 2013, \mnras, 433, 2857 

\bibitem[Meiksin \& White(2003)]{2003MNRAS.342.1205M} Meiksin, A., \& White, M.\ 2003, \mnras, 342, 1205 

\bibitem[M{\'e}nard et al.(2011)]{2011MNRAS.417..801M} M{\'e}nard, B., Wild, V., Nestor, D., et al.\ 2011, \mnras, 417, 801 

\bibitem[M{\'e}nard et al.(2013)]{2013arXiv1303.4722M} M{\'e}nard, B., Scranton, R., Schmidt, S., et al.\ 2013, arXiv:1303.4722 

\bibitem[Meurer et al.(1999)]{1999ApJ...521...64M} Meurer, G.~R., Heckman, T.~M., \& Calzetti, D.\ 1999, \apj, 521, 64 

\bibitem[Milliard et al.(2007)]{2007ApJS..173..494M} Milliard, B., Heinis, S., Blaizot, J., et al.\ 2007, \apjs, 173, 494 

\bibitem[Morrison et al.(2017)]{2017MNRAS.467.3576M} Morrison, C.~B., Hildebrandt, H., Schmidt, S.~J., et al.\ 2017, \mnras, 467, 3576

\bibitem[Morrissey et al.(2005)]{2005ApJ...619L...7M} Morrissey, P., Schiminovich, D., Barlow, T.~A., et al.\ 2005, \apjl, 619, L7 

\bibitem[Morrissey et al.(2007)]{2007ApJS..173..682M} Morrissey, P., Conrow, T., Barlow, T.~A., et al.\ 2007, \apjs, 173, 682 

\bibitem[Murthy(2014a)]{2014ApJS..213...32M} Murthy, J.\ 2014, \apjs, 213, 32 

\bibitem[Murthy(2014b)]{2014Ap&SS.349..165M} Murthy, J.\ 2014, \apss, 349, 165 

\bibitem[Newman(2008)]{2008ApJ...684...88N} Newman, J.~A.\ 2008, \apj, 684, 88-101 

\bibitem[Osterbrock \& Ferland(2006)]{2006agna.book.....O} Osterbrock, D.~E., \& Ferland, G.~J.\ 2006, Astrophysics of gaseous nebulae and active galactic nuclei, 2nd.~ed.~by D.E.~Osterbrock and G.J.~Ferland.~Sausalito, CA: University Science Books, 2006, 

\bibitem[Ouchi et al.(2008)]{2008ApJS..176..301O} Ouchi, M., Shimasaku, K., Akiyama, M., et al.\ 2008, \apjs, 176, 301 

\bibitem[Oyarz{\'u}n et al.(2017)]{2017ApJ...843..133O} Oyarz{\'u}n, G.~A., Blanc, G.~A., Gonz{\'a}lez, V., Mateo, M., \& Bailey, J.~I., III 2017, \apj, 843, 133 

\bibitem[Paresce \& Jakobsen(1980)]{1980Natur.288..119P} Paresce, F., \& Jakobsen, P.\ 1980, \nat, 288, 119 

\bibitem[P{\^a}ris et al.(2018)]{2018A&A...613A..51P} P{\^a}ris, I., Petitjean, P., Aubourg, {\'E}., et al.\ 2018, \aap, 613, A51 

\bibitem[Patej \& Eisenstein(2016)]{2016MNRAS.460.1310P} Patej, A., \& Eisenstein, D.\ 2016, \mnras, 460, 1310 

\bibitem[Planck Collaboration et al.(2014)]{2014A&A...571A..16P} Planck Collaboration, Ade, P.~A.~R., Aghanim, N., et al.\ 2014, \aap, 571, A16 

\bibitem[Pullen et al.(2014)]{2014ApJ...786..111P} Pullen, A.~R., Dor{\'e}, O., \& Bock, J.\ 2014, \apj, 786, 111 

\bibitem[Pullen et al.(2018)]{2018MNRAS.478.1911P} Pullen, A.~R., Serra, P., Chang, T.-C., Dor{\'e}, O., \& Ho, S.\ 2018, \mnras, 478, 1911 

\bibitem[Rahman et al.(2015)]{2015MNRAS.447.3500R} Rahman, M., M{\'e}nard, B., Scranton, R., Schmidt, S.~J., \& Morrison, C.~B.\ 2015, \mnras, 447, 3500 

\bibitem[Rahman et al.(2016a)]{2016MNRAS.457.3912R} Rahman, M., M{\'e}nard, B., \& Scranton, R.\ 2016, \mnras, 457, 3912 

\bibitem[Rahman et al.(2016b)]{2016MNRAS.460..163R} Rahman, M., Mendez, A.~J., M{\'e}nard, B., et al.\ 2016, \mnras, 460, 163 

\bibitem[Reddy et al.(2018)]{2018ApJ...853...56R} Reddy, N.~A., Oesch, P.~A., Bouwens, R.~J., et al.\ 2018, \apj, 853, 56 

\bibitem[Reid et al.(2016)]{2016MNRAS.455.1553R} Reid, B., Ho, S., Padmanabhan, N., et al.\ 2016, \mnras, 455, 1553 

\bibitem[Salpeter(1955)]{1955ApJ...121..161S} Salpeter, E.~E.\ 1955, \apj, 121, 161 

\bibitem[Schiminovich et al.(2005)]{2005ApJ...619L..47S} Schiminovich, D., Ilbert, O., Arnouts, S., et al.\ 2005, \apjl, 619, L47 

\bibitem[Schlafly \& Finkbeiner(2011)]{2011ApJ...737..103S} Schlafly, E.~F., \& Finkbeiner, D.~P.\ 2011, \apj, 737, 103 

\bibitem[Schlegel et al.(1998)]{1998ApJ...500..525S} Schlegel, D.~J., Finkbeiner, D.~P., \& Davis, M.\ 1998, \apj, 500, 525 

\bibitem[Schmidt et al.(2013)]{2013MNRAS.431.3307S} Schmidt, S.~J., M{\'e}nard, B., Scranton, R., Morrison, C., \& McBride, C.~K.\ 2013, \mnras, 431, 3307 

\bibitem[Schmidt et al.(2015)]{2015MNRAS.446.2696S} Schmidt, S.~J., M{\'e}nard, B., Scranton, R., et al.\ 2015, \mnras, 446, 2696 

\bibitem[Scottez et al.(2016)]{2016MNRAS.462.1683S} Scottez, V., Mellier, Y., Granett, B.~R., et al.\ 2016, \mnras, 462, 1683

\bibitem[Seljak \& Warren(2004)]{2004MNRAS.355..129S} Seljak, U., \& Warren, M.~S.\ 2004, \mnras, 355, 129 

\bibitem[Silva et al.(2013)]{2013ApJ...763..132S} Silva, M.~B., Santos, M.~G., Gong, Y., Cooray, A., \& Bock, J.\ 2013, \apj, 763, 132 

\bibitem[Skibba et al.(2014)]{2014ApJ...784..128S} Skibba, R.~A., Smith, M.~S.~M., Coil, A.~L., et al.\ 2014, \apj, 784, 128 

\bibitem[Slosar et al.(2011)]{2011JCAP...09..001S} Slosar, A., Font-Ribera, A., Pieri, M.~M., et al.\ 2011, JCAP, 9, 001 

\bibitem[Sobral et al.(2013)]{2013MNRAS.428.1128S} Sobral, D., Smail, I., Best, P.~N., et al.\ 2013, \mnras, 428, 1128 

\bibitem[Tinker et al.(2010)]{2010ApJ...724..878T} Tinker, J.~L., Robertson, B.~E., Kravtsov, A.~V., et al.\ 2010, \apj, 724, 878 

\bibitem[Trenti \& Stiavelli(2008)]{2008ApJ...676..767T} Trenti, M., \& Stiavelli, M.\ 2008, \apj, 676, 767 

\bibitem[Vanden Berk et al.(2001)]{2001AJ....122..549V} Vanden Berk, D.~E., Richards, G.~T., Bauer, A., et al.\ 2001, \aj, 122, 549 

\bibitem[Werk et al.(2014)]{2014ApJ...792....8W} Werk, J.~K., Prochaska, J.~X., Tumlinson, J., et al.\ 2014, \apj, 792, 8 

\bibitem[Wold et al.(2014)]{2014ApJ...783..119W} Wold, I.~G.~B., Barger, A.~J., \& Cowie, L.~L.\ 2014, \apj, 783, 119 

\bibitem[Wold et al.(2017)]{2017ApJ...848..108W} Wold, I.~G.~B., Finkelstein, S.~L., Barger, A.~J., Cowie, L.~L., \& Rosenwasser, B.\ 2017, \apj, 848, 108 

\bibitem[Xu et al.(2005)]{2005ApJ...619L..11X} Xu, C.~K., Donas, J., Arnouts, S., et al.\ 2005, \apjl, 619, L11 

\bibitem[Zehavi et al.(2011)]{2011ApJ...736...59Z} Zehavi, I., Zheng, Z., Weinberg, D.~H., et al.\ 2011, \apj, 736, 59 

\end{thebibliography}

\appendix

\section{Parameterization of the UVB emissivity}\label{App:parameterization}

In Section~\ref{subsubsec:spectral_tagging_Galex} we apply the spectral tagging technique to constrain the spectrum of the UVB using a simple parameterization visualized in Figure~\ref{fig:CUB_illustration}. This corresponds to a rest-frame comoving volume emissivity 

\begin{equation}
   \epsilon_{\nu}(\nu, z)=
   \begin{cases}
     \epsilon_{1500}\, \Bigg(\dfrac{\nu}{\nu_{1500}}\Bigg)^{\alpha_{1500}}  &   \,\,\,\,\,\,\,\,\,\,\,\,\,\,\,\,\,\,\,\,\text{if } \Bigg(\dfrac{c}{\nu}\Bigg) > 1216\,\AA \;; \\\\
     \epsilon_{1500}\, \Bigg(\dfrac{\nu_{1216}}{\nu_{1500}}\Bigg)^{\alpha_{1500}}\Bigg[ \Bigg(\dfrac{\nu}{\nu_{1216}}\Bigg)^{\alpha_{1100}} + \textrm{EW}_{\rm Ly\alpha}\,\dfrac{\nu^2}{c}\, \delta_{\rm D}(\nu-\nu_{1216}) \Bigg]  &  \,\,\,\,\,\,\,\,\,\,\,\,\,\,\,\,\,\,\,\,\text{if }1216\,\AA>\Bigg(\dfrac{c}{\nu}\Bigg) > 912\,\AA \;; \\\\
     f_{\rm LyC}\,\epsilon_{1500}\, \Bigg(\dfrac{\nu_{1216}}{\nu_{1500}}\Bigg)^{\alpha_{1500}}\,\Bigg(\dfrac{\nu_{912}}{\nu_{1216}}\Bigg)^{\alpha_{1100}}\,\Bigg(\dfrac{\nu}{\nu_{912}}\Bigg)^{\alpha_{900}} &  \,\,\,\,\,\,\,\,\,\,\,\,\,\,\,\,\,\,\,\,\text{if }\Bigg(\dfrac{c}{\nu}\Bigg) < 912\,\AA \;,
   \end{cases}
\end{equation}

where $\nu_{x} = c/x$ with $x$ being a wavelength label in units of $\AA$, $\epsilon_{1500}$ is the continuum emissivity normalization at $1500\;\AA$, $\textrm{EW}_{\rm Ly\alpha}$ is the $\rm Ly\alpha$ line equivalent width, $\delta_{D}$ is the Dirac delta function for the line shape, and $f_{\rm LyC}$ is the ionizing Lyman continuum escape fraction. We fix the slope of the Lyman continuum $\alpha_{900} = -1.5$ independent of redshift following  \cite{1992ApJ...389L...1M}. This has no effect on our emissivity inference as the Lyman continuum is not detected. For all the other power indices, emissivity normalization, and line- and break-strength parameters, we allow them to evolve with redshift each with one additional parameter using simple functional forms as shown in Figure~\ref{fig:parameters_vs_z}. For linear quantities, they follow a power law of $(1+z)$; for already logarithmic quantities, they are allowed to scale with $\textrm{log}\,(1+z)$. For $\epsilon_{1500}$, $\alpha_{1500}$, and $\alpha_{1100}$, we normalize them at $z=0$:
\begin{eqnarray}
&\epsilon_{1500} = \epsilon_{1500}^{z=0}\,(1+z)^{\gamma_{\epsilon 1500}}\,;\nonumber\\
&\alpha_{1500} = \alpha_{1500}^{z=0}+ C_{\alpha 1500}\,\textrm{log}\,(1+z)\;;\nonumber\\
&\alpha_{1100} = \alpha_{1100}^{z=0}+ C_{\alpha 1100}\,\textrm{log}\,(1+z)\;,
\label{eq:par_as_fn_of_z_a}
\end{eqnarray}
where $\gamma_{\epsilon 1500}$, $C_{\alpha 1500}$, and $C_{\alpha 1100}$ are the redshift evolution parameters for each. For Ly$\alpha$, the direct constraint is at $z\approx 0.3$ and $z\approx1$ when the line is in FUV and NUV bands, respectively (middle panel in Figure~\ref{fig:CUB_illustration}). We therefore parameterize its redshift evolution pivoted at these two redshifts: 
\begin{eqnarray}
\textrm{EW}_{\rm Ly\alpha} &=& C_{\rm Ly\alpha}\,\textrm{log}\,\Bigg(\cfrac{1+z}{1+0.3}\Bigg)+ \textrm{EW}_{\rm Ly\alpha}^{z=0.3}\;;\nonumber\\
\textrm{where } C_{\rm Ly\alpha} &=& (\textrm{EW}_{\rm Ly\alpha}^{z=1}-\textrm{EW}_{\rm Ly\alpha}^{z=0.3}) / \textrm{log}\,\Bigg(\cfrac{1+1}{1+0.3}\Bigg)\;.
\label{eq:par_as_fn_of_z_b}
\end{eqnarray}
This is simply a linear function allowing the equivalent width to be positive (emission) or negative (absorption), and also allows the change of sign over redshift (fourth panel in Figure~\ref{fig:parameters_vs_z}). The Lyman continuum escape fraction is defined to be a positive, logarithmic parameter. The GALEX data will provide direct constraints on the ionizing photons only at $z\approx1$ in FUV and $z\approx2$ in NUV. We therefore have
\begin{eqnarray}
\textrm{log}\,f_{\rm LyC} &=& C_{\rm LyC}\,\textrm{log}\,\Bigg(\cfrac{1+z}{1+1}\Bigg)+ \textrm{log}\,f_{\rm LyC}^{z=2}\;;\nonumber\\
\textrm{where } C_{\rm LyC} &=& (\textrm{log}\,f_{\rm LyC}^{z=2}-\textrm{log}\,f_{\rm LyC}^{z=1}) / \textrm{log}\,\Bigg(\cfrac{1+2}{1+1}\Bigg)\;.
\label{eq:par_as_fn_of_z_c}
\end{eqnarray}

\section{Covariance of the UVB parameters}\label{App:Covariance}

\begin{figure*}[t]
    \begin{center}
         \includegraphics[width=1.\textwidth]{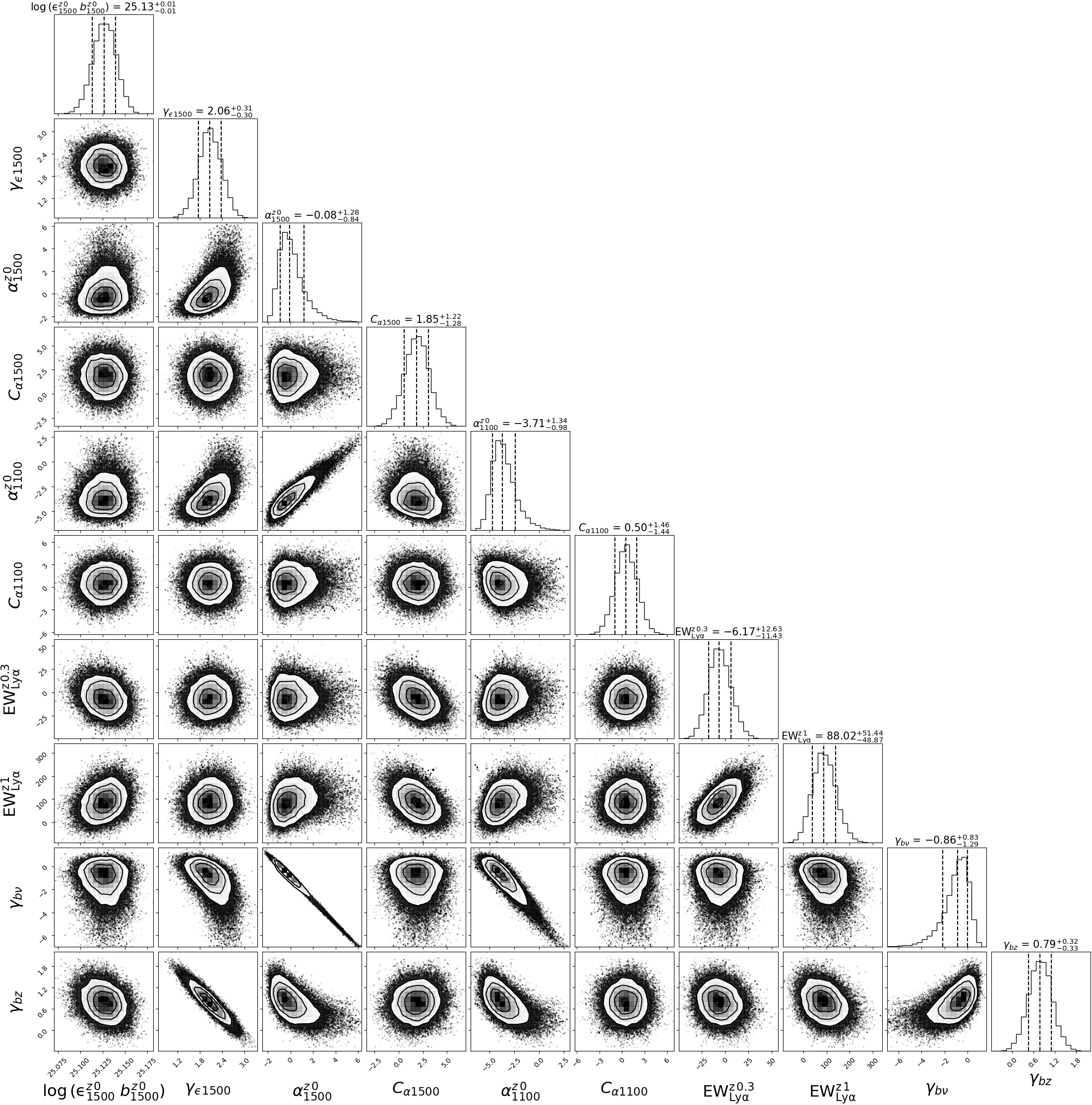}
    \end{center}
    \caption{Triangle plot of the posterior probability distribution for parameters in our UVB model. Diagonal panels show the marginalized posterior for each parameter, the other panels show the projected correlations between each combination of parameter pairs. The normalization $\epsilon_{1500}^{z=0}\;b_{1500}^{z=0}$ has units of $\rm erg\; s^{-1}\; Hz^{-1}\; Mpc^{-3}$.}
    \label{fig:corner}
\end{figure*}

In Section~\ref{subsubsec:MCMC} we use an MCMC method to sample the posteriors of our parameterized UVB emissivity and photon bias given the data, with the best-fit parameters summarized in Table~\ref{table:priors}. Here in Figure~\ref{fig:corner}, we visualize the marginalized posterior distribution for each parameter and their covariances using a triangle plot \citep[using the corner package from][]{corner}. We exclude $\rm{log}\,f_{\rm LyC}^{z=1}$ and $\rm{log}\,f_{\rm LyC}^{z=2}$ in this figure as the ionizing Lyman continuum at both $z=1$ (constrained in FUV) and $z=2$ (constrained in NUV) are not detected and almost entirely independent from other parameters (see Figure~\ref{fig:model_grid}). One can see significant covariances between some of the parameters. For example, the redshift dependence of the $1500\,\AA$ emissivity normalization $\gamma_{\epsilon 1500}$ is degenerate with that of the clustering bias $\gamma_{bz}$, with only their product tightly constrained by the data. A similar degeneracy can be seen for the $1500\,\AA$ spectral slope $\alpha_{1500}^{z=0}$ and the frequency dependence of the clustering bias $\gamma_{b\nu}$.

\end{document}